\documentclass[%
 reprint,
 amsmath,amssymb,
 aps,
]{revtex4-2}

\usepackage{graphicx}
\usepackage{dcolumn}
\usepackage{bm}
\usepackage{epstopdf}
\usepackage{xcolor}
\usepackage{soul}
\usepackage{hyperref}

\makeindex
\makeindex

\begin{document}

\preprint{APS/123-QED}

\title{Realistic Threat Models for Fiber and Free-Space \\ Continuous-Variable Quantum Key Distribution}

\author{Zhiyue Zuo}
\affiliation{School of Automation, Central South University, Changsha 410083, China}
\author{Masoud Ghalaii}
\affiliation{Department of Computing and Mathematics, Manchester Metropolitan University, Manchester M1 5GD, United Kingdom}
\author{Stefano Pirandola}
\affiliation{Department of Computer Science, University of York, York YO10 5GH, United Kingdom}

             
\begin{abstract}
Future global quantum communication networks, or quantum Internet, will realize high-rate secure communication and entanglement distribution for large-scale users over long distances. Continuous
variable (CV) quantum key distribution (QKD) provides a powerful setting for secure quantum communications, thanks to the use of room-temperature off-the-shelf optical devices and the potential
to reach high rates. However, the achievable performance of CV-QKD protocols is fundamentally limited by the fact that they appear to be fragile to both loss and noise. In this study, we provide
a general framework for analyzing the composable finite-size security of CV-QKD with Gaussian-modulated coherent-state protocol (GMCS) under various levels of trust for the loss and noise
experienced by the users of the protocol. Our work is comprehensive of several practical scenarios, encompassing both active and passive eavesdropping configurations, with both wired (i.e.,
fiber-based) and wireless (i.e., free-space and satellite-based) quantum communication channels. Our numerical results evaluate the robustness of the GMCS protocol under varying levels of trust
and demonstrate that it is difficult for a practical protocol to remain robust against untrusted loss at the transmitter. In the wireless case, we analyze a scenario with a sun-synchronous satellite, showing that its key distribution rate, even with the worst level of trust, can outperform a ground chain of ideal quantum repeaters. Our results indicate that—when it comes to engineering and optimizing quantum-safe networks—it is essential to mitigate the shortcomings caused by critical trade-offs between rate performance, trust level, system noise, and communication distance.
\end{abstract}

\maketitle


\section{Introduction}
Secure data transmission over shared network infrastructure is of fundamental importance to the real world. 
Currently, this is achieved through key distribution protocols, among others. 
However, such 'classical' protocols rely on computational intractability and, therefore, cannot create information-theoretic security~\cite{renner2008security}. A promising solution to this problem is quantum key distribution (QKD)~\cite{pirandola2020advances}, which uses the laws of quantum physics to establish information-theoretic security over insecure channels.
In detail, quantum physics promises that measurements on two correlated quantum systems can yield identical outcomes that are fundamentally unpredictable to any eavesdropper. 

Over the past four decades, considerable attention has been paid to QKD \cite{xu2020secure}.
With several metropolitan-scale demonstration systems already in place and on-chip technique development progressing at a steady pace \cite{zhang2019integrated}, the time is right to begin transitioning to a global quantum communications network \cite{harney2022analytical}.
A complete quantum network should, in particular, integrate both fiber and free-space links, and should operate under both night-time and day-time conditions for a large number of users.
For example, in 2021, we saw the implementation of an integrated space-to-ground quantum communication network over 4600 km based on discrete-variable (DV) QKD protocol, in which information is encoded on the polarization of a photon and extracted by single-photon detector at the receiver side~\cite{chen2021integrated}. 

Unlike the DV counterpart, CV-QKD protocols distribute key information using the quadrature components of quantum light for encoding and cost-effective homodyne \cite{mountogiannakis2022composably} or heterodyne \cite{mountogiannakis2022data} for detection, being already compatible with the current global-sacle telecommunication network at room temperature \cite{weedbrook2012gaussian}.
Continuous-variable QKD can potentially encode many secret bits per quantum system and are very suitable for the short-range network with high-rate implementations \cite{pirandola2022architectures,mele2025maximum}. 
In the view of quantum information theory, it can approach the ultimate rate limits of quantum communications, as represented by the repeaterless Pirandola-Laurenza-Ottaviani-Banchi (PLOB) bound~\cite{pirandola2017fundamental}.
On top of that, from an implementation perspective, thanks to the local oscillator (LO) acting as a noise filter, only background noise that is mode-matched with the LO will be detected in CV-QKD \cite{pirandola2021limits}, ensuring its feasibility for daytime operation. 
Moreover, the LO also carries frequency reference information, which can compensate for the space-time curvature effect in satellite-based network \cite{bruschi2014spacetime}.

To date, the most advanced CV-QKD protocols are the Gaussian-modulated coherent-state (GMCS) protocols \cite{grosshans2002continuous}, which benefit from the most well-established security proofs \cite{pirandola2020advances}. In contrast, the security analysis of other CV protocols, such as discrete-modulation CV-QKD, is still under active development \cite{leverrier2009unconditional,liu2022theoretical}.
We note that security proofs for CV-QKD protocols require some assumptions to make the information-theoretic security valid.
In particular, the two legal users, namely, Alice (sender) and Bob (receiver), hope to have isolated or trusted devices, such that they are inaccessible to the eavesdropper Eve.
Therefore, most current security proofs are based on the ideal models of the quantum devices, or equivalently, assuming the practical devices behave as ideal models during the protocol operation.

Unfortunately, in its practical form, CV-QKD devices are prone to implementation side channels, like all modern information systems. 
In the worst-case scenario, all the noise present in the channel or setup may be accessible to Eve for obtaining information \cite{pirandola2021composable}.
One idea to avoid attacks that exploit device imperfections is to use device-independent (DI) approaches \cite{marshall2014device}, which do not require both Alice’s and Bob’s devices to be trusted. 
Such protocols are immune from many side-channel attacks, but offer significantly lower key rates than protocols with trusted devices. 
Moreover, DI CV-QKD needs to add non-Gaussian resources to violate a Bell inequality, which is difficult to prepare~\cite{paternostro2009violations}.
A partial version of DI protocols is the measurement-device-independent QKD protocols \cite{pirandola2015high,fletcher2025overview}, which removes threats from the detector’s point of view, but still assumes that the state-preparation devices of Alice and Bob are completely trusted and has a lower key rate. 

Motivated by the above facts, before designing a practical quantum network, it is necessary to develop a realistic threat model for CV-QKD protocols, which take the devices' imperfection into account, so that one can investigate the practical ultimate limit, or achievable rate, within a network. 
In this paper, we propose a realistic threat model for one-way GMCS protocols, using the local local-oscillator (LLO) scheme \cite{qi2015generating}, under both fiber and free-space links.
Starting from a discussion over various devices' imperfections, we investigate the composable security of GMCS protocol, which allows specifying the security requirements for combining different cryptographic applications in a unified and systematic way \cite{muller2009composability}. 
In the fiber-based case, we consider a scenario where the quantum channel is multiplexed with classical channels. This can reduce the high deployment and maintenance costs of fiber resources.
We then start with the considerations of the noise from the classical system and then determine the main noise in the LLO scheme. 
In the free-space case, we consider the satellite-to-ground communication channel with an advanced downlink setting, where the quasntum satellite operates on a low Earth orbit (LEO). We then show how a satellite can beat the fiber-based repeater-based bounds, which shows its potentiality for a high-rate global quantum network.
Unlike previous studies that focused solely on noise trust levels~\cite{usenko2016trusted,weedbrook2010quantum,weedbrook2012continuous,nag2025continuous}, we present a unified and comprehensive framework that accounts for all trust levels of both noise and loss on Alice’s and Bob’s sides.

The rest of this paper is organized as follows.
In Sec.~\ref{sec:sys} and Sec.~\ref{sec:3}, we provide a general framework for the composable security of CV-QKD, which accounts for levels of trust in the loss and noise of the communication.
In Sec.~\ref{sec:fiber}, we investigate the setup noise of the LLO scheme added at both the transmitter and receiver.
In Sec.~\ref{sec:fib}, we consider fiber-based quantum communications multiplexed with classical systems.
In Sec.~\ref{sec:sat}, we extend to the satellite-to-ground case, showing that suitably high key rates can be achieved under various trust levels.
Finally, Sec.~\ref{sec:con} is for conclusions.

\section{System description} \label{sec:sys}
\label{sec:systemdesc}

\subsection{General theoretical model} 
\label{sec:Notation} 
\begin{figure}
\centerline{\includegraphics[width=3.6in]{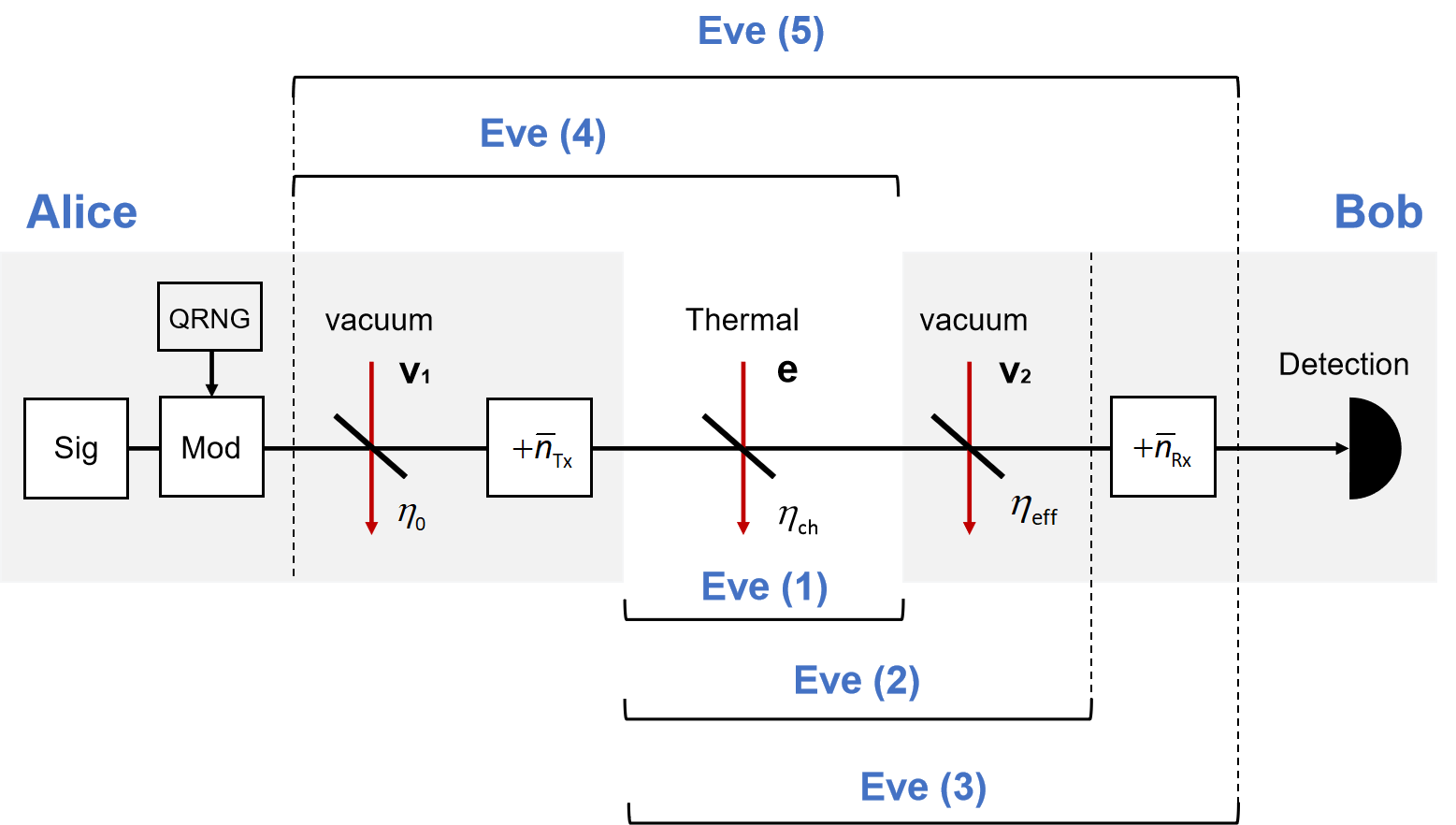}}
	\caption{\label{fig:theory} Quantum communication scenario between transmitter (Alice) and receiver (Bob) separated by a quantum channel with transmittance $ \eta_{\rm ch} $ and mean number of input thermal photons $ {\bar n_e} $.
Alice’s setup has extra thermal photons $ {\bar n}_{\rm Tx} $.
Bob’s setup has detection efficiency $ \eta_{\rm eff} $ and extra thermal photons $ {\bar n}_{\rm Rx} $.
Sig: signal laser;
Mod: random Gaussian displacement;
QRNG: quantum random-number generator.
We also describe the various trust levels. 
Eve (1): Eve only attacks the quantum channel. 
Eve (2): Eve collects
leakage from both the quantum channel and the receiver’s setup.
Eve (3): besides the two leakages above, Eve also performs an active side-channel attack on Bob's setup, so that the noise internal to the setup is untrusted.
Eve (4) and Eve (5) are enhanced versions of Eve (1) and Eve (3), respectively.
In these cases, Eve can collect leakage from Alice's setup and control the Alice's setup noise.
}
\end{figure}

In Fig.~\ref{fig:theory},  we show the  theoretical model of the one-way GMSC CV-QKD protocol.
The sender Alice performs a random Gaussian displacement operation, in terms of the zero-mean and $\mu_{0}-1$ variance random data generated by the quantum random-number generator (QRNG) \cite{gehring2021homodyne}, on a set of vacuum states for preparing the initial coherent state $ \left| \alpha_0  \right\rangle $, whose amplitude can be decomposed as $\alpha_0=(q_0+ip_0)/2$.
Here we represent QRNG's classical inputs, i.e., values of either $q_0$ or $p_0$, by a generic notation $x_0$. Then, the total quadrature variance including vacuum after the displacement is $\mu_0$.
Followed by the Gaussian displacement, a variable optical attenuator (VOA), with transmittance $\eta_{0}$ and modeled as a beam splitter (BS), attenuates $ \left| \alpha_0  \right\rangle $ to get Alice's transmitted coherent state $ \left| \alpha  \right\rangle $ with the amplitude $\alpha=(q+ip)/2$.
In detail, one should set a suitable $\eta_{0}$ to make the variance of $x=q$ or $p$ match the modulation variance expected by the protocol, assumed to be $ \sigma_x^2=\mu-1$.
Thus, the suitable transmittance of VOA should be $ {\eta_0} = {{\sigma_x^2} \mathord{\left/
 {\vphantom {{\sigma_x^2} {\left( {{\mu_0} - 1} \right)}}} \right.
 \kern-\nulldelimiterspace} {\left( {{\mu_0} - 1} \right)}} $, and the total quadrature variance including vacuum after the VOA becomes $\mu$. In what follows, we represent the quadratures by a generic notation $\hat x=\hat q, \hat p$ with the canonical commutation relations $ [\hat q, \hat p]=2i$ \cite{weedbrook2012gaussian} .
Then, the generic quadrature of Alice's transmitted signal can be written as  
\begin{align}
\hat x = x + \sqrt {{\eta_0}} {\hat x_0}+ \sqrt {1 - {\eta_0}} {\hat x_{{v_1}}} + {z_{{\rm{Tx}}}},
\end{align}
where ${\hat x_0}$ is the quadrature operator associated with the vacuum noise of the input bosonic mode, ${\hat x_{v_1}}$ denotes the quadrature of the vacuum mode $v_1$, and $x:=\sqrt{\eta_0}x_0$ (with $x_0$ being the classical random variable of the QRNG).
Here, we have accounted for the imperfections of Alice's devices that introduce various extra thermal noise, with mean photon number ${\bar n}_{\rm Tx}$, before transmission. This is described by the Gaussian variable ${z_{{\rm{Tx}}}}$, which has variance $ 2{\bar n}_{\rm Tx}$.

Next, the attenuated signal undergoes a thermal-loss channel, with transmittance $ \eta_{\rm ch} $ and mean number of thermal noise photons $ {\bar n_e}$, that is controlled by Eve.
Thus, the thermal photon added to the input signal via this channel has mean value $ {\bar n_{\rm ch}}={\bar n_e} (1-\eta_{\rm ch}) $.
At the receiver side, Bob’s setup is characterized by detection efficiency $ \eta_{\rm eff} $ and the setup noises caused by the imperfections of Bob's devices, whose mean photon number is $ {\bar n}_{\rm Rx} $.
Bob performs homodyne or heterodyne detection for his classical output $y$, and with the aid of classical data post-processing, Alice and Bob share a sequence of secure bits \cite{jain2022practical}. 
The input-output relation for the total channel from the classical input $x$ to the output $y$ takes the form
\begin{align}\label{e:cla}
y = \sqrt{\tau} x + z,
\end{align}
where $\tau = \eta_{\rm eff} \eta_{\rm ch} $, and $z$ is a noise variable given by
\begin{equation}
\begin{aligned}
z &= \sqrt {{\tau_0}} {\hat x_0} + \sqrt {\tau \left( {1 - {\eta_{\rm{0}}}} \right)} {\hat x_{{v_1}}} +   \sqrt {{\eta_{{\rm{eff}}}}\left( {1 - {\eta_{{\rm{ch}}}}} \right)} {\hat x_e} \\
&+ \sqrt {1 - {\eta_{{\rm{eff}}}}} {\hat x_{{v_2}}} + \sqrt \tau  {z_{{\rm{Tx}}}}+{z_{{\rm{Rx}}}} + {z_{\det }}\label{eq3},
\end{aligned}
\end{equation}
where $\tau_0= \eta_{\rm eff} \eta_{\rm ch} \eta_{0} $, ${\hat x_{e}}$ denotes the quadrature of the thermal mode $e$ with variance $2{\bar n_e}+1$, ${\hat x_{v_2}}$ is the quadrature of the vacuum mode $v_2$, and ${z_{\det }}$ is an additional Gaussian variable with variance $v_{\rm det}-1$. Here, the detector shot-noise has $v_{\rm det}=1$ and $v_{\rm det}=2$ for the homodyne and heterodyne detector, respectively. In detail, `1' accounts for fundamental vacuum fluctuations, and $v_{\rm det}-1$ is the additional penalty that depends on the specific type of final detection: 0 for homodyne and 1 for heterodyne.
Therefore, the variance of $z$ can be written as
\begin{align}
\sigma_z^2 = 2\bar n + {v_{\det }},
\end{align}
where $\bar n$ is the mean number of total thermal photons due to the channel noise and setup noise.
In terms of the analysis above, $\bar n$ can be decomposed as 
\begin{align}\label{e:noise}
\bar n = \tau {\bar n_{{\rm{Tx}}}} + {\eta_{{\rm{eff}}}}{\bar n_{{\rm{ch}}}} + {\bar n_{{\rm{Rx}}}}.
\end{align}
Finally, the total signal-to-noise ratio (SNR) is given by
\begin{align}\label{e:SNR}
{\rm{SNR = }}\frac{{\tau \sigma_x^2}}{{\sigma_z^2}} = \frac{{\sigma_x^2}}{\Upsilon},
\end{align}
where $ \Upsilon = \sigma_z^2/\tau $ is the equivalent noise at Alice's side. 
Note that one can also use excess noise, in shot noise units, to represent the noise instead of the mean number of thermal photons.

\subsection{Adversarial behaviors} 

Let us describe the trust level of CV-QKD, which depends on the assumption about Eve's attack capability.
Once we have clarified the side channel and the noise sources in the protocol, we can identify various trust levels in terms of different assumptions for Eve. 
As shown in Fig.~\ref{fig:theory}, Eve (1) denotes the basic attack scenario, where Eve only attacks the quantum channel by inserting her photons and collecting all the photons in the side-channel port.
In this scenario, both Alice's and Bob's setup are inaccessible to Eve so the loss and noise on both sides are considered trusted.
However, if Eve can access Bob, she may monitor or control the receiver’s setup for her attack.
For example, the Eve (2) scenario means Eve collects leakage from both the quantum channel and the receiver’s setup.
The photon leakages of the receiver are stored by Eve and become part of her attack. 
In a worse scenario, e.g., Eve (3), besides the two types of leakages, Eve can also actively tamper with the receiver, so that the noise internal to the setup is considered untrusted.
Similarly, Eve may monitor or control Alice’s setup, as shown in the Eve (4) or Eve (5) scenario.
The worst scenario is Eve (5) where no loss and noise in both Alice's and Bob's setup is trusted. 
This scenario, clearly, has the lowest secret key rate than previous cases because all the leakages and noises are under Eve’s control so the protocol needs higher security.
Note that a well-known trade-off of QKD is that increasing security decreases the rate, i.e., the rate-security trade-off \cite{pirandola2022architectures}.

Another aspect related to Eve's attack capability is the attack model \cite{portmann2022security}.
Here we consider the collective Gaussian attack, where Eve perturbs the channel in an independent and identical way while storing all her outputs in a quantum memory (to be optimally measured at the end of the protocol). 
Two main characteristics of this attack are important for the security of the GMSC protocol. 
First, the collective Gaussian attack has proven to be the most dangerous collective attack \cite{navascues2006optimality, garcia2006unconditional}.
Therefore, in the view of classical variables, the channel between Alice and Bob can be assumed to be linear with additive Gaussian noise, i.e., Eq.~\eqref{e:cla}.
Second, the collective attacks are as efficient as the coherent attacks assuming the permutation symmetry of the classical postprocessing \cite{renner2009finetti}.
Note that the coherent attacks, where Eve is only restricted by by laws of quantum mechanics, are considered the most powerful attacks.
The most general description of the collective Gaussian attack is that all the untrusted leakages are represented by a purification of the environmental BS, while all the untrusted noise photons are considered part of a two-mode squeezed vacuum (TMSV) state in Eve’s hands \cite{pirandola2008characterization, pirandola2021composable}.
An example is the Fig.~\ref{fig:Eve5} in Sec.~\ref{sec:ele}.

\section{Composable finite-size key rate analysis}\label{sec:3}

At the end of the QKD protocol, Alice and Bob share an amount of perfectly correlated information unknown to Eve, which we assume results in $n$ bits of secret key.
Then, these secret bits can be used in many data applications, e.g. data encryption. 
However, not all the results of detection can be used as secret bits.
On the one hand, Alice and Bob need to publicize a random subset of signals for parameter estimation (PE) of the channel, whose knowledge is crucial for applying the most appropriate procedures of error correction (EC) and privacy amplification (PA).
On the other hand, a realistic CV-QKD implementation needs to consider the imperfection in EC and PA, which may abort the protocol. 

In this section, we first show the asymptotic key rate under the collective Gaussian attack.
Then, we take the imperfection of data post-processing into account for a composable finite-size key rate.
The security analysis of all trust levels is based on the prepare-measure representation of the protocol without switching to an entanglement-based (EB) version.
Note that most previous works use the equivalent EB scheme instead of the prepare-measure scheme for the convenience of security analysis.
However, the EB scheme assumes Eve holds the purification of the system between Alice and Bob, which means no losses and noises in both Alice's and Bob's setup are trusted i.e., Eve (5).
Therefore, we use the prepare-measure scheme in the following for multiple trust levels.

\subsection{Asymptotic key rate} 

Let us compute the asymptotic key rate that the users would be able to achieve if they could use the quantum channel for an infinite round.
The asymptotic key rate $ R_{\rm asy}$ follows the Devetak-Winter theorem~\cite{devetak2005distillation}, which equals the gap between the mutual information of the trusted users (scaled by the efficiency of the reconciliation process, $\beta$) and Eve’s Holevo bound, i.e., 
\begin{align}\label{e:R}
{R_{{\rm{asy}}}} = \beta {I_{{\rm{AB}}}}\left( {x:y} \right) - \chi_{\rho},
\end{align}
where $\beta$ denotes the reconciliation efficiency, ${I_{{\rm{AB}}}}\left( {x:y} \right)$ is the Shanon mutual information between Alice and Bob, $ \chi_{\rho} $ is the Holevo bound with a single-copy (Gaussian) state $\rho$, representing the maximum information Eve can steal per use of the channel.
In detail, ${I_{{\rm{AB}}}}\left( {x:y} \right)$ is related to the SNR in Eq.~\eqref{e:SNR} given by
\begin{align}\label{e:I}
{I_{{\rm{AB}}}}\left( {x:y} \right) = \frac{{{v_{\det }}}}{2}{\log_2}\left( {1 + {\rm{SNR}}} \right).
\end{align}
In fact, ${I_{{\rm{AB}}}}\left( {x:y} \right)$ is the channel capacity of the additive white Gaussian noise channel.

The Holevo bound $ \chi_{\rho} $ shows a difference when the protocol uses direct reconciliation (where Bob infers $x$ from $y$) and reverse reconciliation (where Alice infers $y$ from $x$).
In our study, reverse reconciliation is used as it can beat the 3 dB transmission limit of standard direct reconciliation \cite{grosshans2002reverse}.
In reverse reconciliation, $\chi_\rho {\left( {y:{\rm{E}}} \right) }$ is related to the joint covariance matrix between Eve and Bob, which has the form
\begin{align}\label{e:BEE''}
{{\rm{V}}_{BEE''}} = \left( {\begin{array}{*{20}{c}}
{b{\rm{I}}}&C\\
{{C^T}}&{{{\rm{V}}_{EE''}}}
\end{array}} \right),
\end{align}
where $b=\tau \sigma_x^2+2\bar n +1$ is the variance of output before detection, $ \rm{I}=\rm{diag}(1,1) $, $C = \left( {\theta {\rm{I,}}\gamma {\rm{Z}}} \right)$ is the cross-correlation block with $ \rm Z=\rm{diag}(1,-1) $, and ${\rm{V}}_{EE''}$ is Eve’s reduced covariance matrix given by
\begin{align}\label{e:EE''}
{{\rm{V}}_{EE''}} = \left( {\begin{array}{*{20}{c}}
{\phi {\rm I}}&{\psi {\rm Z}}\\
{\psi {\rm Z}}&{\omega {\rm I}}
\end{array}} \right),
\end{align}
where $\omega$ is Eve's the mean number of thermal photons given by $\omega  = 2{\bar n_{{\rm{Eve}}\left( {\rm{i}} \right)}} + 1$, where $\bar n_{\rm Eve (i)}$ denotes the mean number of photons in Eve’s modes. 
The elements of $ {{\rm{V}}_{EE''}} $ and $C$ change with the trust level of protocol, as shown in Sec.~\ref{sec:ele}.
Once these elements are determined, $ \chi_\rho {\left( {y:{\rm{E}}} \right) } $ is given by \cite{weedbrook2012gaussian}
\begin{align}\label{e:Eve}
\chi_\rho {\left( {y:{\rm{E}}} \right) } = \sum\limits_{k = 1,2} {H({v_k})}  - \sum\limits_{k = 3,4} {H({v_k})},
\end{align}
where the $H$-function is defined as
\begin{align}
H(\nu ): = \frac{{\nu  + 1}}{2}{\log_2}\frac{{\nu  + 1}}{2} - \frac{{\nu  - 1}}{2}{\log_2}\frac{{\nu  - 1}}{2}.
\end{align}
The first entropy term in Eq.~\eqref{e:Eve} can be computed from the symplectic spectrum $ \left\{ {{v_1},{v_2}} \right\} $ of $ {{\rm{V}}_{EE''}} $, while the second entropy term is computed from the symplectic spectrum $ \left\{ {{v_3},{v_4}} \right\} $ of Eve’s conditional covariance matrix on Bob’s outcome $ {{\rm{V}}_{\left. {EE''} \right|B}} $.
For homodyne detector, $ {{\rm{V}}_{\left. {EE''} \right|B}} $ follows 
\begin{align}
{{\rm{V}}_{\left. {EE''} \right|B}} = {{\rm{V}}_{EE''}} - {b^{ - 1}}{C^{\rm{T}}}\Pi C,
\end{align}
where $\Pi:=\rm{diag}(1,0) $.
For heterodyne detector, $ {{\rm{V}}_{\left. {EE'} \right|B}} $ follows 
\begin{align}
{{\rm{V}}_{\left. {EE''} \right|B}} = {{\rm{V}}_{EE''}} - {(b + 1)^{ - 1}}{C^{\rm T}}C.
\end{align}

\subsection{The elements of $C$ and $ {{\rm{V}}_{EE''}} $ under various trust levels}\label{sec:ele}

\begin{figure}
\centerline{\includegraphics[width=3.4in]{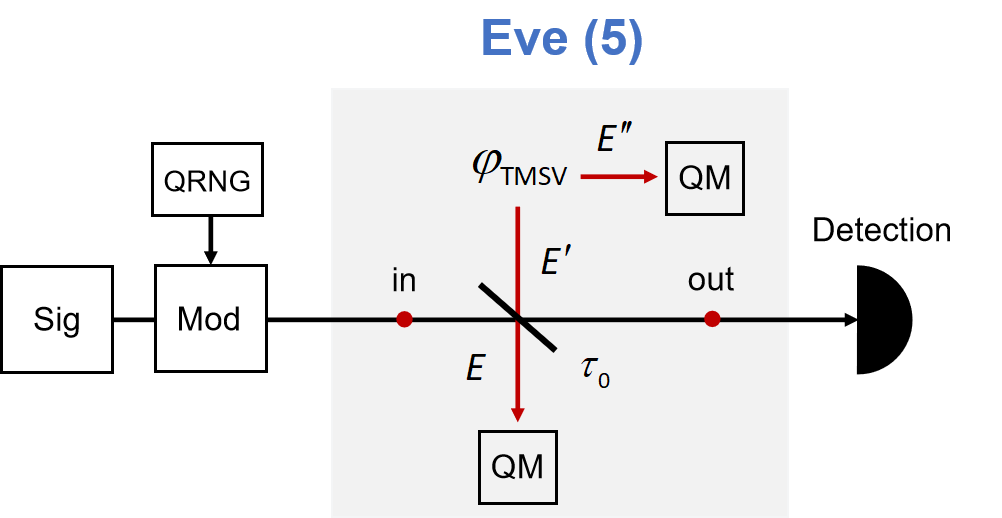}}
\caption{\label{fig:Eve5} The collective Gaussian attack under the assumption of untrusted loss and noise in both Alice and Bob side, i.e., Eve (5) in Fig.~\ref{fig:theory}. 
QM: quantum memory;
${\varphi _{{\rm{TMSV}}}}$: TMSV state.
}
\end{figure}

The elements of $ C $ and $ {{\rm{V}}_{EE''}} $ are related to the trust level of the protocol.
Here we consider the derivation of Eve (5) for example, and the case of other trust levels can be found by corresponding modifications. 
Fig.~\ref{fig:Eve5} shows Eve’s collective Gaussian attack under Eve (5).
In detail, all loss and noise are controlled by Eve so that all the setup noises are considered to come from Eve's TMSV state via the BS with total transmittance $\tau_0$.
Therefore, the variance of Eve's modes (i.e., mode $E'$ and mode $E''$) is given by
\begin{align}
\omega  = 2{\bar n_{{\rm{Eve}}{\kern 1pt} (5)}} + 1 = 2\frac{{\bar n}}{{1 - {\tau_0}}} + 1.
\end{align}
Before the BS, the quadrature $ {\hat x_{{\rm{in}}}} $ is given by ${\hat x_{{\rm{in}}}} = x_0 + {\hat x_0}$ with variance $\mu_0$.
When $ {\hat x_{{\rm{in}}}} $ and $ {\hat x_{E'}} $ pass through BS, the new quadrature $ {\hat x_{{\rm{out}}}} $ and $ {\hat x_{E}} $ have the form
\begin{equation}\label{e:BS}
\left[ {\begin{array}{*{20}{c}}
{{{\hat x}_{{\rm{out}}}}}\\
{{{\hat x}_E}}
\end{array}} \right] = \left[ {\begin{array}{*{20}{c}}
{\sqrt {{\tau_0}} }&{\sqrt {1 - {\tau_0}} }\\
{ - \sqrt {1 - {\tau_0}} }&{\sqrt {{\tau_0}} }
\end{array}} \right]\left[ {\begin{array}{*{20}{c}}
{{{\hat x}_{{\rm{in}}}}}\\
{{{\hat x}_{E'}}}
\end{array}} \right].
\end{equation}
Note that $ {\hat x_{{\rm{out}}}} $ has a variance $b$, while $ {\hat x_{E}} $ is stored by Eve.
Based on Eq.~\eqref{e:BS}, the remaining elements of $ C $ and $ {{\rm{V}}_{EE''}} $ are given by
\begin{equation}
\phi {\rm{ = }}\left\langle {\hat x_E^2} \right\rangle  = {\tau_0}\omega  + \left( {1 - {\tau_0}} \right){\mu_{\rm{0}}},
\end{equation}
\begin{equation}
\psi {\rm{ = }}\left\langle {{{\hat x}_E}{{\hat x}_{E''}}} \right\rangle  = \sqrt {{\tau_0}\left( {{\omega ^2} - 1} \right)},
\end{equation}
\begin{equation}
\theta {\rm{ = }}\left\langle {{{\hat x}_{{\rm{out}}}}{{\hat x}_E}} \right\rangle  = \sqrt {{\tau_0}\left( {1 - {\tau_0}} \right)} \left( {\omega  - {\mu_0}} \right),
\end{equation}
\begin{equation}
\gamma {\rm{ = }}\left\langle {{{\hat x}_{{\rm{out}}}}{{\hat x}_{E''}}} \right\rangle  = \sqrt {\left( {1 - {\tau_0}} \right)\left( {{\omega ^2} - 1} \right)}.
\end{equation}

Then, with a similar derivation, one can get the elements of other trust levels, as shown in Appendix \ref{app:CM}.
The key step is to modify Eve's TMSV state and its input channel.
For Eve (1) scenario, Eve only controls the quantum channel so that Eve's modes have a mean number of photons ${{\bar n}_{{\rm{Eve}}(1)}}{\rm{ = }}{{\bar n}_e}$ and then input the channel with the transmittance $1-\eta_{\rm ch}$.
For Eve (2) with untrusted-loss and trusted-noise detector, the mean number of photons in Eve's modes becomes ${\bar n_{{\rm{Eve}}({\rm{2}})}}{\rm{ = }}{{{\eta_{{\rm{eff}}}}{{\bar n}_{{\rm{ch}}}}} \mathord{\left/
 {\vphantom {{{\eta_{{\rm{eff}}}}{{\bar n}_{{\rm{ch}}}}} {\left( {1 - \tau } \right)}}} \right.
 \kern-\nulldelimiterspace} {\left( {1 - \tau } \right)}}$, while its input transmittance is modified to $1-\tau$.
Next, when the loss and noise of the detector become untrusted i.e., Eve (3), the mean number of photons in mode $E'$ and  mode $E''$ increases to ${\bar n_{{\rm{Eve}}({\rm{3}})}}{\rm{ = }}{{\left( {{\eta_{{\rm{eff}}}}{{\bar n}_{{\rm{ch}}}}{\rm{ + }}{{\bar n}_{{\rm{Rx}}}}} \right)} \mathord{\left/
 {\vphantom {{\left( {{\eta_{{\rm{eff}}}}{{\bar n}_{{\rm{ch}}}}{\rm{ + }}{{\bar n}_{{\rm{Rx}}}}} \right)} {\left( {1 - \tau } \right)}}} \right.
 \kern-\nulldelimiterspace} {\left( {1 - \tau } \right)}}$ with the same input transmittance as Eve (2).
Finally, the Eve (4) scenario has an untrusted Alice so that the mean number of photons in Eve's mode need take Alice's setup noise into account and given by ${\bar n_{{\rm{Eve}}(4)}} = {{\left( {{{\bar n}_{{\rm{ch}}}} + {\eta_{{\rm{ch}}}}{{\bar n}_{{\rm{Tx}}}}} \right)} \mathord{\left/
 {\vphantom {{\left( {{{\bar n}_{{\rm{ch}}}} + {\eta_{{\rm{ch}}}}{{\bar n}_{{\rm{Tx}}}}} \right)} {\left( {1 - {\eta_{\rm{0}}}{\eta_{{\rm{ch}}}}} \right)}}} \right.
 \kern-\nulldelimiterspace} {\left( {1 - {\eta_{\rm{0}}}{\eta_{{\rm{ch}}}}} \right)}}$, whose input transmittance is modified to $1-{\eta_{\rm{0}}}{\eta_{{\rm{ch}}}}$.
Note that the above derivation shows that it is not difficult to decouple the trusted setup noise from the untrusted channel noise by looking at the covariance matrix.
The reason is that one can assume Gaussian channels here as Eve’s optimal attacks for Gaussian modulation schemes correspond to Gaussian channels.
When the modulation changes to discrete schemes, the Gaussian attacks are not expected to be optimal and, thus, making the decoupling of trusted noise difficult \cite{lin2020trusted}.

\subsection{Parameter estimation}\label{sec:PE}

The asymptotic key rate can be calculated once Alice and Bob know the value of all the parameters entered in Eq.~\eqref{e:R}.
However, in the real world, the users may not know all the parameters but have $n_{\rm pe}$ unknown parameters.
Therefore, they need to randomly choose and publicly disclose part of the distributed signals for PE \cite{portmann2022security}.
Because realistic users can only use the quantum channel a finite number of rounds $N$, this estimation is not perfect, which decreases the rate.
In general, Alice and Bob choose and disclose $m$ of the $N$ rounds for the estimation of $\tau$ and $ \bar n $, i.e., $n_{\rm pe}=2$.
Note that it is acceptable to assume that Alice knows the signal modulation $\mu$ and transmittance $\eta_{0}$, while Bob monitors the quantum efficiency $\eta_{\rm eff}$.
Moreover, the PE's data size $m_{\rm pe}$ equals $m$ and $2m$ for the homodyne detector and heterodyne detector, respectively.
With the disclosed pairs $ \left\{ {{x_i},{y_i}} \right\}_{i = 1}^{{m_{{\rm{pe}}}}} $, the users construct the unbiased estimators  \cite{pirandola2021composable}
\begin{align}
\hat \tau  = {\left( {\frac{{\sum\limits_{i = 1}^{{m_{{\rm{pe}}}}} {{x_i}{y_i}} }}{{\sum\limits_{i = 1}^{{m_{{\rm{pe}}}}} {x_i^2} }}} \right)^2},
\end{align}
\begin{align}
\widehat {\bar n} =  - \frac{{{v_{\det }}}}{2} + \frac{1}{{2{m_{{\rm{pe}}}}}}\sum\limits_{i = 1}^{{m_{{\rm{pe}}}}} {{{\left( {{y_i} - \sqrt {\hat \tau } {x_i}} \right)}^2}}.
\end{align}
Here $ \left\{ {{x_i},{y_i}} \right\}_{i = 1}^{{m_{{\rm{pe}}}}} $ are Gaussian as well as independent and identically distributed since Eve performs a collective Gaussian attack.
By assuming a certain number $w$ of confidence intervals, the worst-case estimators are given by
\begin{align}\label{e:tau_low}
\tau ' \simeq \tau  - 2w\sqrt {\frac{{2{\tau ^{\rm{2}}} + \tau \sigma_z^2/\sigma_x^2}}{{{m_{{\rm{pe}}}}}}},
\end{align}
\begin{align}\label{e:n_up}
\bar n'' \simeq \bar n + w\frac{{2\bar n + {v_{\det }}}}{{\sqrt {2{m_{{\rm{pe}}}}} }}.
\end{align}
The confidence parameter $w$ is related to the acceptable error probability of the estimator ${\varepsilon_{\rm{pe}}}$, which can be expressed as~\footnote{When ${{\varepsilon _{{\rm{pe}}}} > {{10}^{ - 17}}}$, we can use ${\sqrt 2 {\rm{er}}{{\rm{f}}^{ - 1}}\left( {1 - 2{\varepsilon _{{\rm{pe}}}}} \right)}$ for $w$. However, if ${{\varepsilon _{{\rm{pe}}}} \le  {{10}^{ - 17}}}$, that formula creates divergences so we need to use ${\sqrt {2\ln \left( {1/{\varepsilon _{{\rm{pe}}}}} \right)} }$. In our work, we set ${\varepsilon _{{\rm{pe}}}}{\rm{ = }}{10^{ - 1{\rm{0}}}} > {10^{ - 17}}$ so we use ${\sqrt 2 {\rm{er}}{{\rm{f}}^{ - 1}}\left( {1 - 2{\varepsilon _{{\rm{pe}}}}} \right)}$}
\begin{align}
w = \left\{ {\begin{array}{*{20}{c}}
{\sqrt 2 {\rm{er}}{{\rm{f}}^{ - 1}}\left( {1 - 2{\varepsilon_{{\rm{pe}}}}} \right)}&{{\varepsilon_{{\rm{pe}}}} > {{10}^{ - 17}}}\\
{\sqrt {2\ln \left( {{1 \mathord{\left/
 {\vphantom {1 {{\varepsilon_{{\rm{pe}}}}}}} \right.
 \kern-\nulldelimiterspace} {{\varepsilon_{{\rm{pe}}}}}}} \right)} }&{{\varepsilon_{{\rm{pe}}}} \le {{10}^{ - 17}}}.
\end{array}} \right. 
\end{align}
Note that the whole estimation is wrong as long as one estimator is wrong so that the PE's total error probability is $ {\varepsilon_{{\rm{pe}}}}\left( {1 - {\varepsilon_{{\rm{pe}}}}} \right) + \left( {1 - {\varepsilon_{{\rm{pe}}}}} \right){\varepsilon_{{\rm{pe}}}} + \varepsilon_{{\rm{pe}}}^2 \simeq 2{\varepsilon_{{\rm{pe}}}} $.
By replacing $\tau$ and $ \bar n$ by $\tau '$ and $ \bar n'' $ respectively in Eq.~\eqref{e:R}, one gets the finite-size rate $R_{{\rm{asy}}}^{{\rm{pe}}} = {R_{{\rm{asy}}}}(\tau ',\bar n'')$.

However, the general PE above is not suitable for all trust levels.
The reason is that PE should consider the most powerful Eve so that an additional estimation on the effect of trust noise (not the value of trust noise) is necessary for Eve (1) to Eve (4).
Take Eve (1) as an example, where both Alice's and Bob's setup noises are trusted such that the untrusted photon only comes from the channel with a mean number ${{{\bar n}_{{\rm{ch}}}}}$.
Based on Eq.~\eqref{e:noise}, one needs the best-case estimators of $\tau {\bar n_{{\rm{Tx}}}}$ and ${\bar n_{{\rm{Rx}}}}$ to maximize ${{{\bar n}_{{\rm{ch}}}}}$ for the most powerful Eve. 
In detail, the best-case estimators of $\tau {\bar n_{{\rm{Tx}}}}$ is $ \tau '{\bar n_{{\rm{Tx}}}} $ since Alice's setup noise is independent of $ \tau $.
In contrast, Bob's setup noise is related to $ \tau $ so that its mean photon number has a best-case estimator ${\bar n_{{\rm{Rx}}}^{{\rm{be}}}}$ (see Sec.~\ref{sec:RXnoise} for details).
In Sec.~\ref{sec:RXnoise}, we find that ${\bar n_{{\rm{Rx}}}^{{\rm{be}}}}$ uses the best-case estimator of $\tau$ instead of $\tau'$, which is given by
\begin{align}
\tau '' \simeq \tau  + 2w\sqrt {\frac{{2{\tau ^{\rm{2}}} + \tau \sigma_x^2/\sigma_x^2}}{{{m_{{\rm{pe}}}}}}}.
\end{align}
Therefore, the finite-size rate of Eve (1) becomes $R_{{\rm{asy}}}^{{\rm{pe}}} = {R_{{\rm{asy}}}}(\tau ',\bar n'',\bar n_{{\rm{Rx}}}^{{\rm{be}}})$.
The same result can also be found for both scenarios Eve (2) and Eve (4).
Note that Eve (3) can use the general estimation in Eq.~\eqref{e:tau_low} and Eq.~\eqref{e:n_up} because Bob's setup noise is untrusted and no best-case estimator is needed.
Finally, all the noises and leakages of Eve (5) are untrusted so the general estimation for $\tau$ and $\bar n$ can be used.
Note that all the estimators of $ {{\bar n}_{{\rm{Eve}}{\kern 1pt} }} $ are based on the estimators of $\tau$ and $\bar n$ thus the total error probability of PE is still $  \simeq 2{\varepsilon_{{\rm{pe}}}} $ for all trust levels.

\subsection{Practical composable security} 

After PE, the remaining $n=N-m$ rounds are used for key generation after error correction and privacy amplification.
First, Alice and Bob convert their $n$ rounds classical input and output into $d$-bit string, i.e., $ {x} \to {k} $ and $ {y} \to {l} $.
Under the action of a collective Gaussian attack, the output classical-quantum state of Alice, Bob, and Eve has the tensor-structure form $ {\rho ^{ \otimes n}} $.
In reverse reconciliation, Bob reveals $\rm leak_{\rm ec}$ bits to help Alice reconstruct Bob’s sequence $ {l^n} $ based on its local sequence $ {k^n} $.
To check whether EC is successful, Bob reveals another $ - {\log_2}{\varepsilon_{{\rm{cor}}}} $ bits to Alice for hash comparison, where $ {\varepsilon_{{\rm{cor}}}} $ bounds the probability that the sequences are different even if their hashes coincide \cite{portmann2022security}.
The protocol goes ahead only when these hashes coincide, thus the tensor-structure state $ {\rho ^{ \otimes n}} $ projects into a non i.i.d. state $ {\tilde \rho ^n} $ before PA.
During PA, the users apply a two-way hash function
over $ {\tilde \rho ^n} $ and output the privacy amplified state $ {\bar \rho ^n} $.
The goal of EC and PA is to make $ {\bar \rho ^n} $ approximates the ideal $s_n$-bit private state $ {\rho_{{\rm{id}}}} $.
We direct readers the Appendix G of Ref.~\cite{pirandola2021limits} for the details of the above steps.

Based on the above steps, we first derive the secret key bound with composable security (without PE).
First, the composable secret key length $s_n$ for $n$ coherent state transmissions satisfies the direct leftover hash bound \cite{tomamichel2011leftover}
\begin{align}\label{e:sn}
{s_n} \ge H_{\min }^{{\varepsilon_s}}{\left( {{l^n}|{{\rm{E}}^n}} \right)_{{{\tilde \rho }^n}}} - {\rm{lea}}{{\rm{k}}_{{\rm{ec}}}} + {\log_2}\left( {\varepsilon_h^2{\varepsilon_{{\rm{cor}}}}} \right),
\end{align}
where $ H_{\min }^{{\varepsilon_s}}{\left( {{l^n}|{{\rm{E}}^n}} \right)_{{{\tilde \rho }^n}}} $ is the smooth min-entropy with the smoothing parameter $\varepsilon_s$, and the security parameter $\varepsilon_h$ characterizes the hashing function.
Here, Eq.~\eqref{e:sn} accounts for all leakage bits due to EC-not only those used for Alice’s reconstruction-providing a more precise expression than that used in previous works~\cite{pirandola2021limits,pirandola2021composable,jain2022practical,goncharov2022continuous}.
To replace the smooth-min entropy of $ {\tilde \rho ^n} $ with that of $ {\rho ^{ \otimes n}} $,one can use the following tensor-product reduction~\cite{pirandola2024improved}
\begin{align}\label{e:reduction}
H_{\min }^{^{{\varepsilon_s}}}{\left( {{l^n}|{{\rm{E}}^n}} \right)_{{{\tilde \rho }^n}}} \ge H_{\min }^{^{{\varepsilon_s}}}{\left( {{l^n}|{{\rm{E}}^n}} \right)_{{\rho ^{ \otimes n}}}},
\end{align}
thus 
\begin{align} 
{s_n} \ge H_{\min }^{^{{\varepsilon_s}}}{\left( {{l^n}|{{\rm{E}}^n}} \right)_{{\rho ^{ \otimes n}}}} - {\rm{lea}}{{\rm{k}}_{{\rm{ec}}}} + {\log_2}\left( {\varepsilon_h^2{\varepsilon_{{\rm{cor}}}}} \right).
\end{align}
Note that the tensor-product reduction in Eq.~\eqref {e:reduction} has been improved compared with previous works \cite{pirandola2021limits,pirandola2021composable,jain2022practical,goncharov2022continuous}, which adopted more pessimistic assumptions.
To further bound the smooth-min entropy of $ {\rho ^{ \otimes n}} $, one may apply the asymptotic equipartition property (AEP) \cite{tomamichel2142framework}
\begin{align}\label{e:AEP}
H_{\min }^{{\varepsilon_s}}{\left( {{l^n}|{{\rm{E}}^n}} \right)_{{\rho ^{ \otimes n}}}} \ge nH{\left( {l|{\rm{E}}} \right)_\rho } - \sqrt n {\Delta_{{\rm{aep}}}},
\end{align}
where $H{\left( {l|{\rm{E}}} \right)_\rho }$ is the conditional von-Neumann entropy, and \cite{pirandola2021composable}
\begin{align} 
{\Delta_{{\rm{aep}}}} \simeq 4{\log_2}\left( {\sqrt {{2^d}}  + 2} \right)\sqrt {{{\log }_2}\left( {{2 \mathord{\left/
 {\vphantom {2 {\varepsilon_{\rm{s}}^2}}} \right.
 \kern-\nulldelimiterspace} {\varepsilon_{\rm{s}}^2}}} \right)}.
\end{align}
Finally, using Eq.~\eqref{e:reduction} and Eq.~\eqref{e:AEP} in Eq.~\eqref{e:sn}, one may write the following lower bound
\begin{equation}
\begin{aligned}
{s_n} 
&\ge n\left[ {H\left( l \right) - \chi_\rho {{\left( {l:{\rm{E}}} \right)} } - {n^{ - 1}}{\rm{lea}}{{\rm{k}}_{{\rm{ec}}}}} \right] \\
&- \sqrt n {\Delta_{{\rm{aep}}}} + {\log_2}\left( {\varepsilon_h^2{\varepsilon_{{\rm{cor}}}}} \right) \label{e:sn2},
\end{aligned} 
\end{equation}
where the conditional von-Neumann entropy is decomposed to $ H{\left( {l|{\rm{E}}} \right)_\rho } = H\left( l \right) - \chi_\rho {\left( {l:{\rm{E}}} \right) } $ with the Shannon entropy $ {H\left( l \right)} $ and Eve’s Holevo bound $ {\chi_\rho {{\left( {l:{\rm{E}}} \right)} }} $.
Note that the brackets part of Eq.~\eqref{e:sn2} equals to the asymptotic key rate ${R_{{\rm{asy}}}}$ in Eq.~\eqref{e:R} because $ H\left( l \right) - {n^{ - 1}}{\rm{lea}}{{\rm{k}}_{{\rm{ec}}}} = \beta {I_{{\rm{AB}}}}\left( {x:y} \right) $, while $\chi_\rho {\left( {l:{\rm{E}}} \right) } \le \chi_\rho {\left( {y:{\rm{E}}} \right) }$ \cite{jain2022practical}.
Here the latter inequality is due to $ {y} \to {l} $ is a completely positive and trace-preserving map thus the Holevo bound follows the monotonicity property.
Finally, the composable key rate is given by
\begin{equation}\label{e:sn3}
\begin{aligned}
{R_{{\rm{com}}}} 
&= \frac{{{p_{{\rm{ec}}}}{s_n}}}{N} \\
&\ge \frac{{{p_{{\rm{ec}}}}\left[ {n{R_{{\rm{asy}}}} - \sqrt n {\Delta_{{\rm{aep}}}} + {{\log }_2}\left( {\varepsilon_h^2{\varepsilon_{{\rm{cor}}}}} \right)} \right]}}{N},
\end{aligned}
\end{equation}
where $p_{\rm ec}$ is the success probability of hash comparison in EC, while $1-p_{\rm ec}$ is known as the frame error rate.

By replacing ${{R_{{\rm{asy}}}}}$ with $R_{{\rm{asy}}}^{{\rm{pe}}}$ in Eq.~\eqref{e:sn3}, one can get the final composable finite-size key rate~\cite{pirandola2024improved}
\begin{equation}\label{e:R_pe_com}
R_{{\rm{com}}}^{{\rm{pe}}} \ge \frac{{{p_{{\rm{ec}}}}\left[ {nR_{{\rm{asy}}}^{{\rm{pe}}} - \sqrt n {\Delta_{{\rm{aep}}}} + {{\log }_2}\left( {\varepsilon_h^2{\varepsilon_{{\rm{cor}}}}} \right)} \right]}}{N}.
\end{equation}
As mentioned in Sec.~\ref{sec:PE}, the total error probability of PE is $  \simeq 2{\varepsilon_{{\rm{pe}}}} $ for all trust levels.
Therefore, the total epsilon security for all trust levels is 
\begin{equation}
\varepsilon  = {\varepsilon_{{\rm{cor}}}} + {\varepsilon_s} + {\varepsilon_h} + 2{p_{{\rm{ec}}}}{\varepsilon_{{\rm{pe}}}},
\end{equation}
where PE's error probability is multiplied by ${{p_{{\rm{ec}}}}}$ since PE performs before EC. Note that Eq.~\eqref{e:R_pe_com} is a lower bound to the number of secret random bits that Alice and Bob can extract, where the protocol assumes an optimal PA with $\varepsilon$-security.
It is also important to note that one can get the upper bound of composable finite-size rate with similar steps by replacing Eq.~\eqref{e:sn} with the converse leftover hash bound \cite{tomamichel2011leftover}.
The upper bound means that the rate between Alice and Bob must be compressed below the value of this bound if they want to have $\varepsilon$-security assured.
When $N$ is typically large, the upper bound almost coincides with the lower bound in Eq.~\eqref{e:R_pe_com}.
More detail can be found in Ref.~\cite{pirandola2024improved}.

\section{Contributions to setup noise} 
\label{sec:fiber}

\subsection{Practical optical layout} 
\label{sec:layout}

To obtain a more realistic model, we first determine the optical layout, so as the setup devices.
Generally, QKD can be implemented by two schemes: transmitted local-oscillator (TLO) and LLO scheme.
In the following, we mainly consider the LLO scheme as it is easy to meet the shot-noise-limited demand of the homodyne detector \cite{wang2018pilot}. 
For an LLO scheme, the LO is generated by an independent laser at Bob’s side instead of Alice’s signal laser. Therefore, the local LO can no longer be used as a reliable phase reference due to the fast relative phase fluctuations between the two lasers. 
To establish a reliable phase reference, a pilot is multiplexed with the signal for transmission.
For example, the pilot can multiplex with signal in time by using a delay line \cite{marie2017self}, or in polarization by a polarization beam-combiner \cite{wang2018pilot}.

\begin{figure*}
\vspace{+.1cm}
\includegraphics[scale=0.56]{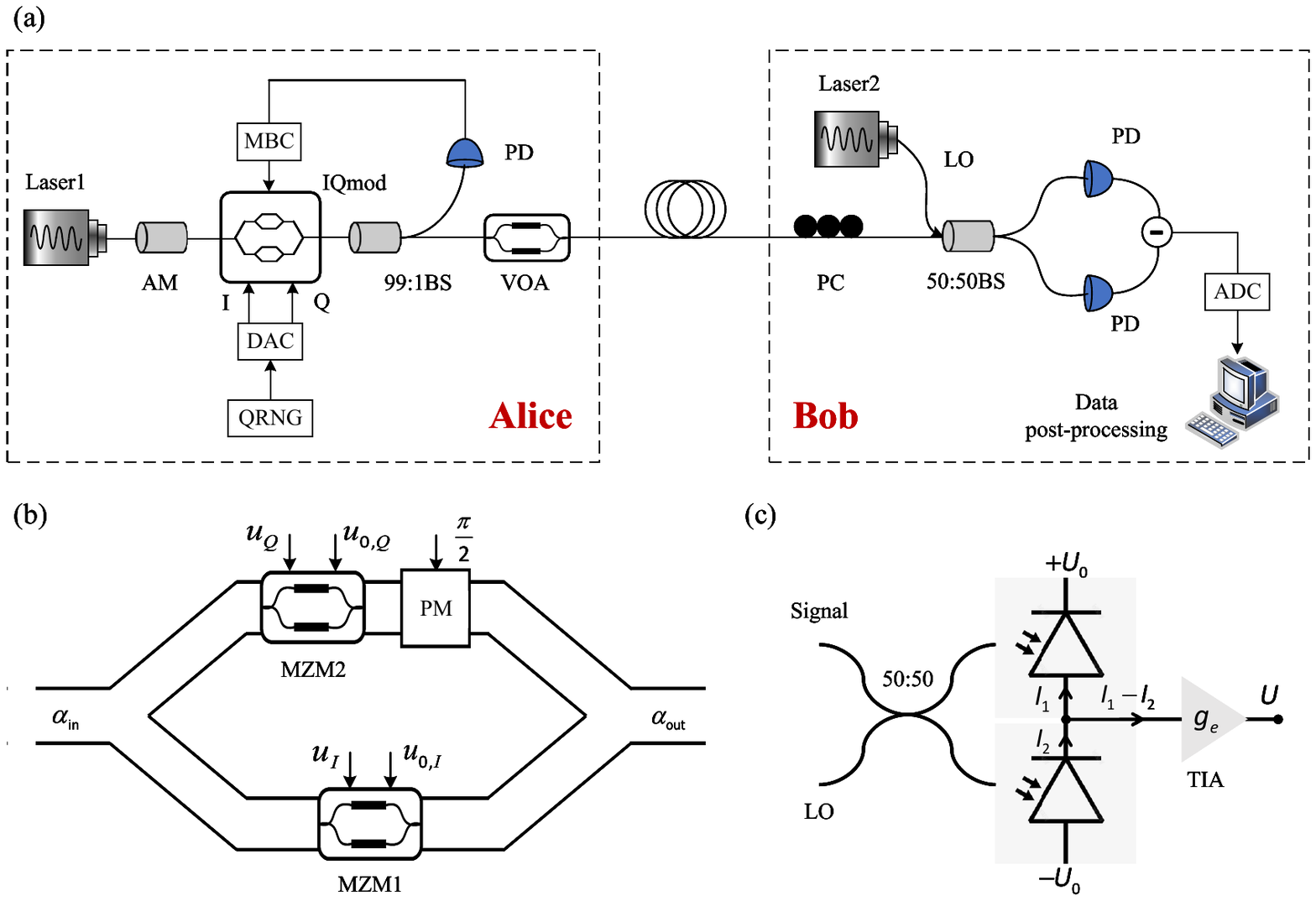}
\caption{\label{fig:system}(a) The optical layout of the LLO scheme.
AM: amplitude modulator;
MBC: modulator bias controller;
DAC: digital-to-analog converter;
IQmod: in-phase and quadrature electro-optic modulator;
PD: photo diode;
BS: beam splitter;
VOA: variable optical attenuator;
PC: polarization controller;
LO: local oscillator; 
ADC: analog-to-digital converter.  
(b) Structure of an optical IQmod.
$ u_I $, $ u_Q $ denote the amplified voltages carrying the information.
$ {u_{0,I}}$ and ${u_{0,Q}} $ are the bias voltage of MZM1 and MZM2, respectively.
MZM: Mach-Zehnder modulator.
PM: phase modulator. 
(c) Schematic of a balanced homodyne detector.
 $+U_0$ and $-U_0$ denote the external linear power supply.
 $ {g_e} $ is the electric amplification, and $U$ is the measurable voltage.
 TIA: transimpedance amplifier.}
\end{figure*}

In this work, we consider the typical pilot-sequential LLO scheme, where the pilot is multiplexed with signal in time, which was first proposed in Ref.~\cite{qi2015generating}.
In detail, the optical layout of this scheme can be simplified as Fig.~\ref{fig:system}(a).
First of all, Laser1 generates the initial signal by amplitude modulator (AM) which is then encoded with Gaussian random data by an in-phase and quadrature modulator (IQmod).
Here the random data is first generated by the QRNG and then performs continuation using a digital-to-analog converter (DAC).
Next, the encoded signal is split by a 99:1 BS, 1 \% of the output is measured by a photo diode (PD), and the result is returned to the modulator bias controller (MBC) to control the direct current (DC) bias voltage of IQmod.
Before the channel, Alice uses a VOA to adjust the modulation variance of the transmitted signal to an expected value.
At Bob's side, a polarization controller (PC) is first used to tune the polarization of the arriving signal.
Then, the signal is mixed with LO (generated by Laser2) by a 50:50 BS followed by two PD for detection.
Finally, the differential output was sampled by an analog-to-digital converter (ADC) for data post-processing.
In what follows, we show the setup noise of this optical layout, where we assume Laser1 is the same as Laser2.

\subsection{Noise added at the transmitter side} 
\label{sec:TXnoise}

\subsubsection{Source noise} 

A real-world laser is not stable all the time but has power fluctuations, mostly due to the fluctuations of the (optical or electrical) pump and the spontaneous emission noise. 
Mathematically, the power fluctuations of a laser are quantified by the relative intensity noise (RIN), which is given by \cite{kleis2022system}
\begin{align}
{\Delta_{{\rm{sig}}}} = \frac{{\left\langle {\delta P{{\left( t \right)}^2}} \right\rangle }}{{{{\left\langle P \right\rangle }^2}}},
\end{align}
where $ \left\langle P \right\rangle $ represents the average power of the laser, and $ {{{(\delta P)}^2}} $ is a time-domain power fluctuation which is assumed to be Gaussian distributed. 
For Laser1, the power fluctuation can be considered as injecting extra thermal photons on the initial quantum signal before modulation. 
More precisely, the fluctuation is equal to injecting extra thermal photons on the transmitted signal with a mean value \cite{laudenbach2018continuous}
\begin{align}\label{e:RIN}
{\bar n_{\rm RIN}}={\bar n_{\rm sig}}\sqrt {{\Delta_{{\rm sig}}}l_{\rm sig}},
\end{align}
where $ l_{\rm sig} $ is the linewidth of signal laser, $ \bar n_{\rm sig}= \sigma_x^2/2 $ represents the (ideal) transmitted signal's mean photon number.
Thus, the mean photon number of the practical signal is given by $ \bar n_{\rm sig}+\bar n_{\rm RIN} $.
As shown in Eq.~\eqref{e:RIN}, a narrower linewidth laser or a laser with low RIN is beneficial for reducing noise.
Besides, this noise will be suppressed if $ {\bar n_{\rm sig}} $, or equivalently, the modulation variance is low.

\subsubsection{Modulation noise}\label{sec:mod}

A typical way for CV-QKD to realize Gaussian modulation is to use the combination of AM and phase modulator (PM)~\cite{wang2018pilot}.
However, IQ modulation has become popular in recent years because it offers a more compact design, cost-effectiveness, and potential robustness against Trojan-horse attacks~\cite{jain2021modulation,jain2014risk}.
Fig.~\ref{fig:system}(b) shows the structure of an optical integrated IQmod, or equivalently, a nested Mach-Zehnder modulator (MZM).
The principle of MZM is to manipulate the refractive index of the material by applying a voltage to a pair of electrodes surrounding the waveguide to achieve modulation.
The reason for using nested structure is because a single MZM can perform amplitude or phase modulation but cannot realize arbitrary trajectories of the complex envelope \cite{kleis2022system}.
In detail, the input field $\alpha_{\rm in}$ is first equally split into two parts. 
Next, the lower part performs amplitude modulation by MZM1 in push-pull mode, while the upper part is also amplitude-modulated by MZM2 and then an additional $\pi/2$ phase shift.
Finally, both signal parts are recombined and output $\alpha_{\rm out}$.
Mathematically, the input-output
relation of an ideal nested MZM is given by
\begin{align}
{\alpha_{{\rm{out}}}} = \frac{{{\alpha_{{\rm{in}}}}}}{2}\left[ {\cos \left( {\frac{{{u_I} + {u_{0,I}}}}{{2{u_\pi }}}\pi } \right) + {\rm{j}}\cos \left( {\frac{{{u_Q} + {u_{0,Q}}}}{{2{u_\pi }}}\pi } \right)} \right],
\end{align}
where $ u_I $ and $ u_Q $ are the amplified voltages carrying the information that is to be transmitted, $ {{u_\pi }} $ is the voltage at which the optical phase changes by $\pi$, $ {u_{0,I}}$ and ${u_{0,Q}} $ are the bias voltage of MZM1 and MZM2, respectively.
We find that the nested MZM is generally nonlinear.

In the modulation process, Alice first uses the DAC to translate the random data into a voltage $ u_{\rm{DAC}} $.
This voltage is then amplified by a factor of $g$ to the voltage required to drive IQmod, i.e., $ u_I $ or $ u_Q $.
However, the imperfection of devices may get a non-ideal voltage $ {u_{{\rm{DAC}}}} \to {u_{{\rm{DAC}}}} + \delta {u} $ with a specific deviation $ \delta u $, which is also amplified by $g$.
Thus, the final drive voltage has an error $ g\delta u $, which adds extra noise.
In addition, the non-ideal bias voltage with deviation $ \delta u_{\rm{DC}} $ also adds noise.
The calculation of voltage deviation noise is shown in Appendix~\ref{app:mod}.
Finally, the transmitted signal is equal to injecting extra thermal photons with a mean value
\begin{align}\label{e:vol}
{\bar n_{{\rm{vol}}}} < {\bar n_{{\rm{sig}}}}{\left[ {\pi \frac{{g\delta u + \delta {u_{{\rm{DC}}}}}}{{2{u_\pi }}} + {{\left( {\pi \frac{{g\delta u + \delta {u_{{\rm{DC}}}}}}{{2\sqrt 2 {u_\pi }}}} \right)}^2}} \right]^2}.
\end{align}
Note that the above equation assumes the bias voltage of PM is perfect without deviation. We find that this noise is also suppressed if $ {\bar n_{\rm sig}} $, or equivalently, the modulation variance is low.

\subsection{Noise added at the receiver side} 
\label{sec:RXnoise}

\subsubsection{Detection noise}

The detector of the GMCS protocol hopes to distinguish the received coherent states with minimum error.
The balanced homodyne detector, which has reached a high degree of maturity and cost efficiency, is considered the simplest setup for such tasks, as it relies only on Gaussian operations~\cite{weedbrook2012gaussian}. 
As shown in Fig.~\ref{fig:system}(c), in a homodyne detector, a 50:50 fused fiber coupler and two positive intrinsic negative PD convert the optical input of signal and LO into electric currents $ I_1 $ and $ I_2 $, respectively.
The current difference $ I_1-I_2 $, which is proportional to the classical output $y$, is then amplified to a measurable voltage $ U $ by a transimpedance amplifie (TIA) given by
\begin{align}\label{e:U}
U{\rm{ = }}{g_e}\left( {{I_1} - {I_2}} \right) = {g_e}{\alpha_{LO}}y,
\end{align}
where $ {g_e} $ is the electric amplification, ${\alpha_{LO}}$ is the amplitude of LO.
However, this measurable voltage may drift in practice due to the imperfections of PD and TIA, thus introducing noise.
In what follows, we consider two noises related to these imperfections: electronic noise and common mode noise.

\paragraph{Electronic noise.}
Electronic noise is related to the dark current of PD, which can be quantified by the noise equivalent power (NEP).
By the way, NEP is also considered the minimum input optical power to generate a photocurrent.
In terms of Ref.~\cite{pirandola2021composable}, the mean value of the extra thermal photos is given by 
\begin{align}
{\bar n_{{\rm{el}}}} = \frac{{{v_{\det }}{\rm{NE}}{{\rm{P}}^2}W\delta t}}{{2hf{P_{\det }}}},
\end{align}
where $W$ is the bandwidth, 
$ {\delta t} $ is the pulse duration, 
$ f $ is the optical frequency for wavelength, 
$ h$ is the Planck's constant, 
and $P_{\det }$ is the detected LO power.
Note that $P_{\det }$ differs for TLO and LLO schemes.
For example, with the same transmit LO power $P_{{\rm{LO}}}$, one has $P_{\det }=
\tau P_{{\rm{LO}}}$ for TLO, and $P_{\det }=P_{{\rm{LO}}}$ for LLO.
In this case, the LLO scheme has a lower electronic noise than that of TLO and its noise does not increase as the transmission distance increases.
In addition, we remark that one usually treats electronic noise as trusted noise in experiments \cite{jain2022practical}.

\paragraph{Common-mode noise.}
A realistic TIA may not only amplify the current difference $ I_1-I_2 $ in Eq.~\eqref{e:U} but add a small portion, whose total measurable voltage can be expressed as
\begin{align}
U = {g_e}({I_1} - {I_2}) + {g_{{\rm{cm}}}}\frac{1}{2}({I_1} + {I_2}). 
\end{align}
Here, $ g_{\rm cm} $ is the common-mode gain, and the so-called common-mode rejection ratio (CMRR) is given by $ {g_{\rm{C}}} = {{{g_e}} \mathord{\left/
 {\vphantom {{{g_e}} {\left| {{g_{{\rm{cm}}}}} \right|}}} \right.
 \kern-\nulldelimiterspace} {\left| {{g_{{\rm{cm}}}}} \right|}} $.
Physically, CMRR reflects the ability of homodyne detector to suppress common-mode noise.
In the ideal case, the common-mode gain will not cause additional noise if the signal and LO have stable power \cite{laudenbach2018continuous}.
However, as discussed above, the realistic power of signal and LO are both fluctuating such that injecting extra thermal photons with a mean value of 
\begin{align}
{\bar n_{\rm{C}}} = \frac{{{v_{\det }}}}{{8g_{\rm{C}}^2}}\left( {\frac{{hf{{\bar n}_{{\rm{sig}}}}}}{{2\delta t{P_{\det }}}} + \frac{{\delta t}}{{hf}}{P_{\det }}} \right){\Delta _{{\rm{sig}}}}{l_{{\rm{sig}}}},
\end{align}
where we set the linewidth and RIN of LO to $ {l_{{\rm{LO}}}} = {l_{{\rm{sig}}}} $ and $ {\Delta_{{\rm{LO}}}} = {\Delta_{{\rm{sig}}}} $ respectively.
Note that the the power fluctuation of the LO laser not only generates the common-mode noise but also directly affects the measurement because the homodyne detector uses LO to amplify the quantum signal [see Eq.~\eqref{e:U}].
Fortunately, the different operation of homodyne detector greatly reduces this effect on quadrature, and as a consequence, the injected thermal photons can be ignored \cite{laudenbach2018continuous}. 

\subsubsection{Quantization noise}

As shown in Fig.~\ref{fig:system}(a), there is an ADC converter after the homodyne detector to map the input analog voltage $U$ into digital code.
However, ADC cannot have a noiseless mapping because the analog voltage is infinite but the digital code is a finite number.
In other words, each digital output can correspond to several input analog voltages.
In detail, a $\bar d$ bits ADC has ${2^{\bar d}}$ voltage intervals so that each interval has a range
\begin{align}
\Delta U = \frac{{\bar U}}{{{2^{\bar d}}}},
\end{align}
where $\bar U$ is the full voltage range.
For example, in the present study we assume a 12-bit ADC with a 1 V full voltage range, which has an interval range $\Delta U$=0.244 mV.
All analog voltage falling into the same interval has the same digital output without distinguishing, which is equal to adding an extract noise on the noiseless mapping.
Totally, the extra thermal photon has a mean value 
\begin{align}\label{e:ADC}
{\bar n_{\rm{Q}}} = \frac{{{v_{{\rm{det}}}}}}{2}\frac{{\delta t}}{{hfg_e^2{\varpi ^2}{P_{\det }}}}\left( {\frac{{{\Delta ^{\rm{2}}}U}}{{12}} + {V_{{\rm{intr}}}}} \right),
\end{align}
where $\varpi$ is the responsivity of PD, and $ {{V_{{\rm{intr}}}}} $ is the intrinsic voltage variance of the ADC.
We find that the quantization noise also shows a difference between TLO and LLO schemes, which comes from a different value of $P_{\det }$ too. In addition, the larger the ADC bit resolution, the lower the quantization noise.
Note that ADC also has thermal noise, which is caused by the physical movement of charge inside electrical conductors.
However, the quantization noise dominates when Bob uses a low-resolution ADC (below 16 bits)~\cite{wang2024high,Texas}.

\subsubsection{Residual phase noise}\label{sec:phase}

The main reason for the phase noise is that the frequency noise of the laser results in a random phase drift of the laser output wave.
Generally, the frequency noise occurs in laser because of the imperfections of the resonator, the temperature fluctuation, and the resonator is not phase selective \cite{kleis2022system}.
Mathematically, the total frequency noise of a laser is characterized in terms of the linewidth, which describes the full width at half maximum (FWHM) of its power spectral density. 
Typically in telecommunications, semiconductor lasers with linewidths of less than 1 MHz are used.

In the LLO scheme, Alice and Bob use two independent lasers, i.e., signal laser and LO laser so that there is a relative phase drift between the two lasers.
This phase drift affects the interference of signal and LO in the homodyne detector, and as a consequence, introduces phase noise.
In other words, the differential phase after the homodyne detector is drifting as well.
To compensate for this phase drift, Alice generates a set of pilot pulses, which are sent to Bob along with the quantum signal.
Then, Bob can detect the pilot, extract phase information, and reduce the phase error.
Even with the aid of pilots, however, the phase error cannot be fully compensated but has a residual error.
A fundamental limit is that the signal and pilot may not have exactly the same phase.
Finally, the mean number of extra thermal photons caused by the residual phase noise can be quantified by
\begin{align}
{{\bar n}_{{\rm{LO}}}} = {{\bar n}_{{\rm{sig}}}}\frac{{2\pi \tau {l_{\rm{W}}}}}{\Theta},
\end{align}
where $ \Theta $ denotes the system clock, ${l_{\rm{W}}}{\rm{ = }}{{\left( {{l_{{\rm{sig}}}}{\rm{ + }}{l_{{\rm{LO}}}}} \right)} \mathord{\left/
 {\vphantom {{\left( {{l_{{\rm{sig}}}}{\rm{ + }}{l_{{\rm{LO}}}}} \right)} {\rm{2}}}} \right.
 \kern-\nulldelimiterspace} {\rm{2}}}$ is the average linewidth. We find that ${\bar n_{{\rm{LO}}}}$ increases with increasing linewidth or modulation variance, which is the reason one prefers narrow-linewidth lasers in LLO experiments.

Note that the phase noise can be reduced in the TLO scheme because the transmitted LO (generated by the same laser as the signal) can be used as a reliable phase reference.
In addition, the phase noise in classical systems can be performed at relatively low penalties because of its discrete modulation formats and high symbol rates (exceeding 10 GBd) \cite{zhou2014modulation}. 
In contrast, in CV-QKD it is extremely challenging to separate the phase noise from the modulated phase due to the ultra low SNR even with a discrete modulation. 
Finally, we remark that another source of the phase noise is the measurement error of the pilot, which is considered as the sole source of phase noise in Ref.~\cite{laudenbach2018continuous}.
In fact, this noise can be suppressed by using a stronger pilot.

\section{fiber-based protocols}\label{sec:fib}

\begin{figure*}
\vspace{+.1cm}
\includegraphics[scale=0.51]{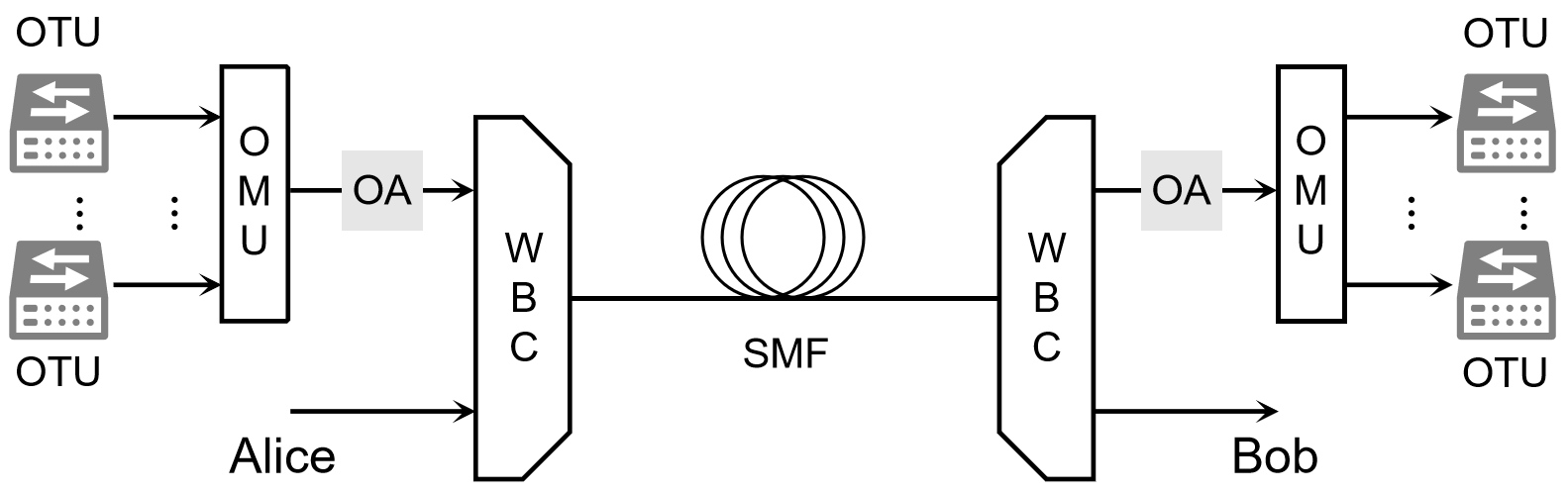}
\caption{\label{fig:DWDM} Fiber-based CV-QKD system multiplexed with classical channels. 
OTU: optical transform unit;
OMU: optical multiplexer unit;
OA: optical amplifier;
WBC: wide band coupler;
SMF: Single-mode fiber.
}
\end{figure*}
 
In this section, we compute the composable finite-size key rate of fiber-based CV-QKD under various trust levels and setup noises.
As shown in Eq.~\eqref{e:noise}, the remaining unknown noise is the thermal photons related to the channel.
In what follows, we first account for the mean value of channel-added thermal photon $ {\bar n_{\rm ch}} $, and then compute the composable finite-size key rates.

In practice, there is an increasing demand for multiplexing of QKD with classical communications in the same fiber because of the high deployment and maintenance costs of fiber resources.
Thus, this section considers that a quantum channel is multiplexed with various classical channels by a wide band coupler (WBC) module, as shown in Fig.~\ref{fig:DWDM}.
In detail, WBC is an all-fiber wideband coupler made using a fused taper method.
Here we set the WBC to a three-wave multiplexer/demultiplexer with wave sections 1310 nm, 1490 nm, and 1530–1565 nm (i.e., the C band)~\cite{geng2021coexistence}.
Then, we select the wave sections 1490 nm and C band for quantum and classical communication respectively.
Finally, the total channel transmittance is then deposited into two parts: WBC transmittance $\eta_{\rm w}$ and fiber transmittance $\eta_{\rm fib}$ i.e., ${\eta_{{\rm{ch}}}} = \eta_{\rm{w}}^{\rm{2}}{\eta_{{\rm{fib}}}}$.
Note that the classical channels may have both forward (Alice to Bob) and backward (Bob to Alice) directions in practice (Fig.~\ref{fig:DWDM} shows the forward direction).

\subsection{Noise added at fiber links} 
\label{sec:channelnoise}

In the following, we discusses the added noise when the quantum signals pass through the standard single-mode fiber (SMF), especially the part that comes from the classical system.
In general, broadband noise photons, including out-of-band photons and in-band photons, may be generated at fiber due to its nonlinear susceptibility, inelastic scattering, and so on.
We will mainly focus on the in-band photons since these photons, mode-matched with the signal, are undesirably superposed on the signal field and disturb the interference.
Note that no impact of nonlinearities and inelastic scattering is expected if only quantum signals are propagated through the fiber, where the amount of induced nonlinear distortion about the signal itself is extremely low \cite{kleis2022system}.

\subsubsection{Nonlinear susceptibility}

Due to the nonlinear susceptibility of silica glass, when two or more pumps exist in the optical fiber, the additional photon frequencies, which are different from those present in the initial fields, can be generated by third-order nonlinearity ${\chi ^{\left( {\rm{3}} \right)}}$.
This nonlinear process is the so-called four-wave mixing (FWM) where no energy is transferred to or taken from the fiber, i.e. no phonon excitation or de-excitation takes place.

For the case that QKD and classical communications are multiplexed, theoretically, the additional frequencies generated by FWM may fall into the wavelength range of the quantum channel, resulting in in-band noise. 
However, the efficient generation of FWM depends on the phase-matching condition, which is easy to fulfill around the zero dispersion wavelength \cite{eraerds2010quantum}.
In our work, the wavelengths of classical communications are located at the C band, where the dispersion coefficient of SMF is around 20 ps${{\rm{nm}}^{ - 1}} {{\rm{km}}^{ - 1}}$.
In contrast, the light with a 1490 nm wavelength is easier to meet the phase-matching condition.
This is one of the reasons why we set the wavelength of the quantum signal instead of the classical signal at 1490 nm.
In addition, the FWM noise can be effectively suppressed by setting sufficient channel separation.
Also, we assume that the quantum channel is 8 THz (40 nm) apart from the nearest classical channel so we simply neglect FWM noise ${\bar n_{{\rm{FWM}}}}$.

\subsubsection{Inelastic scattering}\label{sec:channelnoise.1}

\begin{figure}
	\centerline{\includegraphics[width=3.4in]{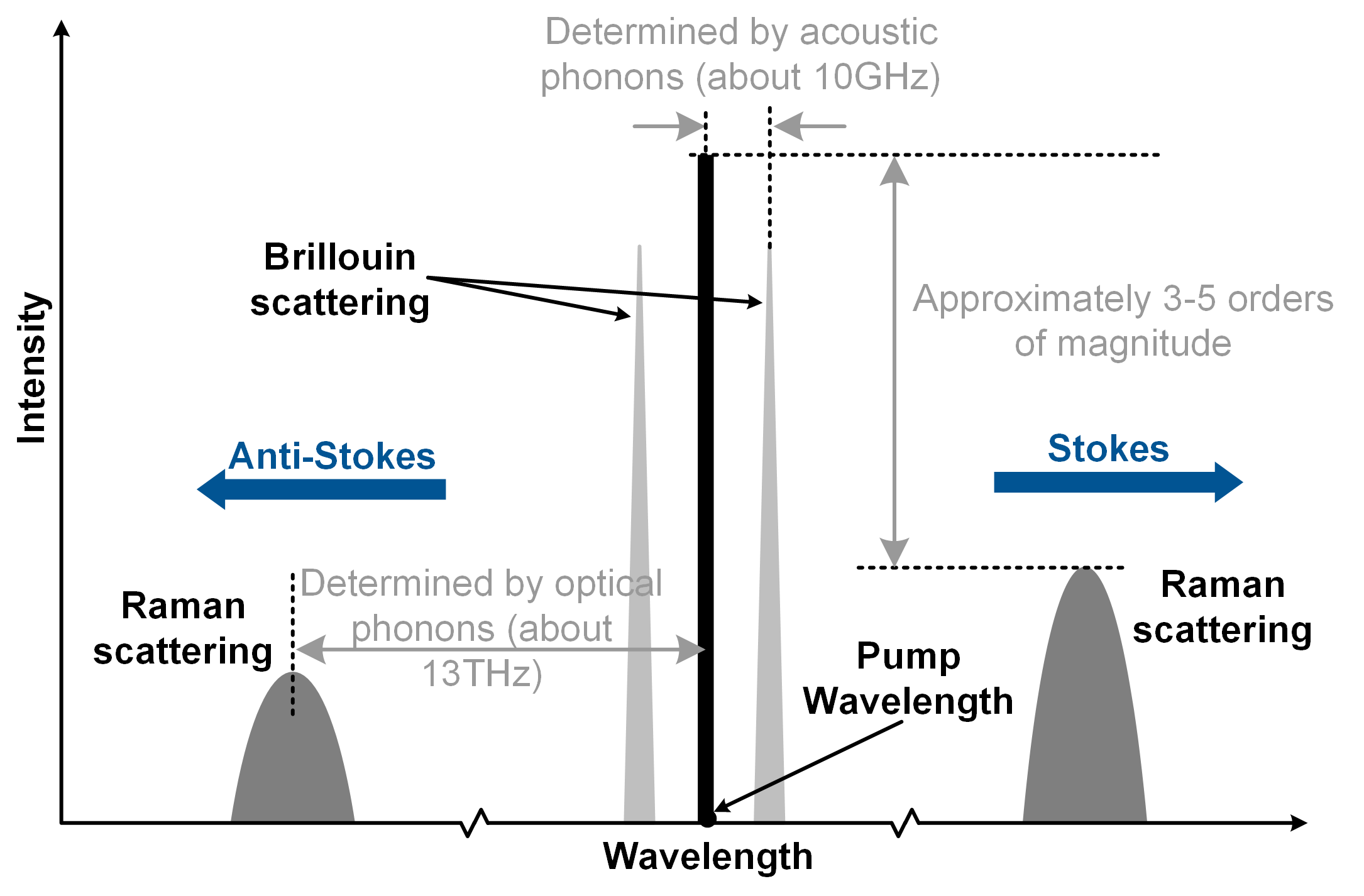}}
	\caption{\label{fig:Ram} Intensity and wavelength distributions of Raman scattering and Brillouin scattering. Here, the pump wavelength is 1550 nm, which is located at C band.}
\end{figure}

Inelastic scattering occurs when the pump exchanges energy with the fiber medium, causing a portion of the pump to be scattered at a different wavelength.
Two different types of scattering fall into this category: Raman scattering and Brillouin scattering.

{\em Raman noise.} 
Raman scattering is an inelastic scattering process in which light interacts and exchanges energy with molecular vibrations of the transport medium. 
In detail, due to photon–phonon interaction in fiber, photons can change their wavelength and thus compromise other channels.
Depending on whether a phonon gets excited or de-exited, photons at frequencies above and below the initial frequencies are generated.
The scattered light components at lower and higher frequencies are referred to as Stokes and anti-Stokes waves respectively.
As shown in Fig.~\ref{fig:Ram}, Raman scattering can lead to significant frequency shifts covering the entire C band, having an intensity maximum at a shift of about 13 THz (corresponding to a wavelength shift of 100 nm at 1550 nm).
In other words, the wavelength of noise photons caused by the classical channels may coincide with that of the quantum signal (i.e., 1490 nm), which cannot be filtered out and will contribute to in-band noise.

Since anti-Stokes scattering is less probable than Stokes scattering \cite{kumar2015coexistence}, we set the classical wavelength in the C band for lower Raman noise, i.e., only anti-Stokes scattering occurs for the 1490 nm quantum signal. Moreover, the anti-Stokes scattering can be further divided into forward scattering (Alice to Bob) and backward scattering (Bob to Alice).
Mathematically, the power of these anti-Stoke Raman noises is given by
\begin{align}\label{e:forward}
{P_{{\rm{Ram}},{\rm{f}}}} = {P_{{\rm{in}}}} \cdot {10^{ - {\bar \alpha} L/10}} \cdot L \cdot {\rho_\lambda } \cdot {\Delta_\lambda },
\end{align}
\begin{align}\label{e:backward}
{P_{{\rm{Ram}},{\rm{b}}}} = {\rm{5}}{P_{{\rm{in}}}} \cdot {\rho_\lambda } \cdot {\Delta_\lambda } \cdot \frac{{1 - {{10}^{ - {\bar \alpha} L/5}}}}{{{\bar \alpha} \ln \left( {10} \right)}},
\end{align}
where  $ {\Delta_\lambda } $ is the filter, 
$ {\bar \alpha} $ is the fiber attenuation coefficient,
$ L $ is the distance, 
${\rho_\lambda }$ is the spontaneous Raman scattering coefficient depending on wavelength $\lambda$, 
$ {P_{{\rm{in}},{\rm{f}}}} $ and $ {P_{{\rm{in}},{\rm{b}}}} $ are the input power of forward classical channel and backward classical channel, respectively.
The detail of the derivation of Eq.~\eqref{e:forward} and Eq.~\eqref{e:backward} is presented in Appendix~\ref{app:Raman}.
In our study, we only need to consider the forward Raman noise because the quadrature detection is located on Bob's side.
In addition, the backward Raman photons, which arrive at Alice's side, can be blocked from the system with an isolator.
Thus, the mean number of Raman noise photons (arriving at the detector) per spatio-temporal and polarization mode is given by
\begin{align}\label{e:Raman}
{\bar n_{{\rm{Ram}}}} = \frac{1}{2} \cdot \frac{c}{{h{f^3}}} \cdot \frac{{{P_{{\rm{Ram}},{\rm{f}}}} \cdot {\eta_{\rm{w}}}}}{{{\Delta_\lambda }}},
\end{align}
where $c$ is the speed of light.
Note that the pre-factor 1/2 is due to the polarization mode selectivity of LO.

\begin{figure}
\vspace{-0.5cm}
\centerline{\includegraphics[width=3.6in]{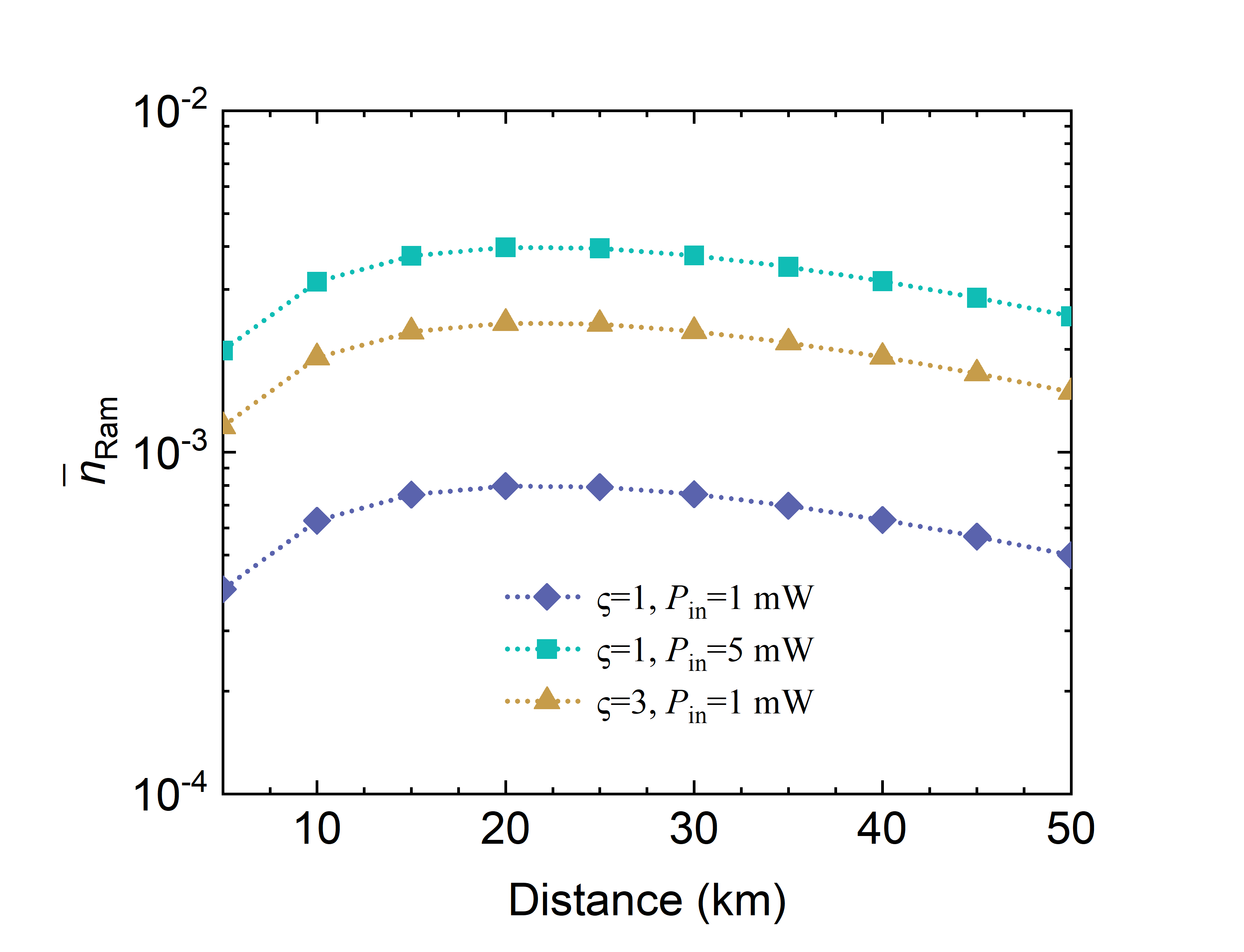}}
	\caption{\label{fig:Raman} The relationship between the mean number of Raman noise photons and transmission distance under various classical power and channel numbers $\varsigma$.}
\end{figure}

Fig.~\ref{fig:Raman} shows the relationship between the mean number of Raman noise photons and transmission distance under various classical power and channel numbers $\varsigma$.
As shown in Ref.~\cite{eraerds2010quantum}, the spontaneous Raman scattering coefficient is wavelength dependent, which is between ${\rm{2}} \times {\rm{1}}{{\rm{0}}^{{\rm{ - 9}}}}$ to ${\rm{4}} \times {\rm{1}}{{\rm{0}}^{{\rm{ - 9}}}}$ ${{\rm{nm}}^{ - 1}} {{\rm{km}}^{ - 1}}$ for C band.
Here we consider the worst case that all the classical channel has ${\rho_\lambda }{\rm{ = 4}} \times {\rm{1}}{{\rm{0}}^{{\rm{ - 9}}}}$ ${{\rm{nm}}^{ - 1}} {{\rm{km}}^{ - 1}}$. 
Therefore, the total Raman noise after multiplexing with multiple classical channels is $\varsigma {\bar n_{{\rm{Ram}}}} $.
In addition, the fiber attenuation coefficient of the C band is set to 0.2 dB/km, while the insertion loss of all wave sections is set to 0.9 (0.46 dB) for simple, which is referred to Ref.~\cite{geng2021coexistence}.
We find that the Raman noise varies with transmission distance and reaches a maximum at 22 km for all cases. This differs from Ref.~\cite{laudenbach2018continuous}, where the Raman noise remains constant across all distances. Note that the Raman noise should depend on the distance, as the Raman photons also experience loss in the channel before arriving at the detector.
Moreover, as shown in Eq.~\eqref{e:Raman}, when fiber parameters and classical system are determined, it is hard to control Raman noise.

{\em Brillouin noise.}
When Brillouin scattering occurs, the pump photon is scattered into a Stokes photon and an optical phonon. 
Similarly, the pump photon can also absorb the energy of a phonon to produce an anti-Stokes photon. 
Like Raman scattering, the frequency of the Stokes photon is lower than that of the pump photon and the frequency of the anti-Stokes photon is higher than that of the pump photon. 
However, in Raman scattering the frequency shift is determined by the optical phonon, while in Brillouin scattering the frequency shift is determined by the acoustic phonon.  
As shown in Fig.~\ref{fig:Ram}, scattering off acoustic phonon is not critical as the maximal frequency shift of the scattered photons is 10 GHz \cite{eraerds2010quantum}, while the 1490 nm quantum channel is 8 THz apart from the nearest classical channel. Therefore, we simply neglect the Brillouin noise ${\bar n_{{\rm{Bri}}}}$.

\subsubsection{Other noises related to the classical system}

In addition to the fiber, other components in classical communication systems may have an impact on quantum signals.
As shown in Fig.~\ref{fig:DWDM}, we usually use an optical amplifier (OA) for classical signals in classical systems, such as an erbium-doped fiber amplifier.
The amplified spontaneous emission (ASE) from a practical OA, which has a broad bandwidth of the order of tens of nm, can be treated as an in-band noise source on Alice's side since the 1490 nm quantum channel is only 40 nm apart from the nearest classical channel.
Fortunately, this noise can be negligible because the WBC used at Alice’s side acts as a bandpass filter and can greatly suppress the ASE noise.
For CV-QKD, the mean number of ASE photons after WBC is given by
\begin{align}
{\bar n_{{\rm{OA}}}} = {\xi_{\rm{w}}}{n_{{\rm{sp}}}}\left( {\vartheta  - 1} \right),
\end{align}
where $ \vartheta  $ is the gain of OA, $ {n_{{\rm{sp}}}} \ge 1 $ is the spontaneous factor, and ${\xi_{\rm w}}$ is the cross-channel isolation of WBC.
As shown in Ref.~\cite{geng2021coexistence}, the value of ${\xi_{\rm w}}$ is high, which can greatly suppress the OA noise.

\subsection{Total amount of noise in fiber-based protocols}

\begin{figure}[b]
	\centerline{\includegraphics[width=3.6in]{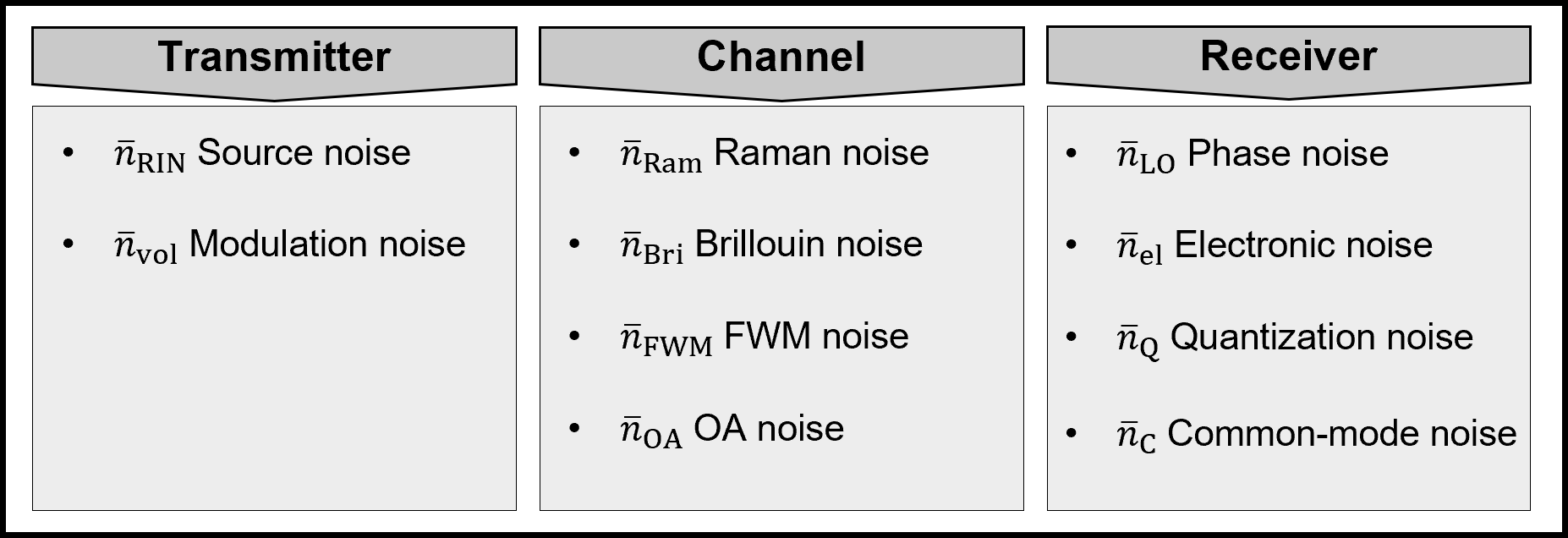}}
	\caption{\label{fig:table:noise} The types of noise considered in Sec.~\ref{sec:fib}.}
\end{figure}

The noises considered in this section are listed in Fig.~\ref{fig:table:noise}.
From Eq.~\eqref{e:noise}, and based on the analysis presented above, one can estimate the total amount of added thermal photons that the receiver sees by assuming
\begin{equation}
{\bar n_{{\rm{Tx}}}} = {\bar n_{{\rm{RIN}}}} + {\bar n_{{\rm{vol}}}},
\end{equation}
\begin{equation}
{\bar n_{{\rm{ch}}}} = {\bar n_{{\rm{Ram}}}},
\end{equation}
\begin{equation}
{\bar n_{{\rm{Rx}}}}{\rm{ = }}{\bar n_{{\rm{el}}}}{\rm{ + }}{\bar n_{\rm{C}}}{\rm{ + }}{\bar n_{{\rm{Q}}}}{\rm{ + }}{\bar n_{{\rm{LO}}}},
\end{equation}
Here we ignore some types of channel noise as discussed above, including the FWM noise ${\bar n_{{\rm{FWM}}}}$, the Brillouin noise ${\bar n_{{\rm{Bri}}}}$, and OA noise ${\bar n_{{\rm{OA}}}}$.
The corresponding expressions in terms of excess noise can be given by 
\begin{align}\label{e:ex}
{\xi_{{\rm{Tx}}}} = 2{\bar n_{{\rm{Tx}}}},\quad {\xi_{{\rm{ch}}}} = \frac{{2{\eta_{{\rm{eff}}}}{{\bar n}_{{\rm{ch}}}}}}{\tau },\quad {\xi_{{\rm{Rx}}}} = \frac{{{{\bar n}_{{\rm{Rx}}}}}}{\tau }.
\end{align}
Note that Eq.~\eqref{e:ex} attributes all excess noise to Alice's side.

\begin{figure}
\vspace{-0.5cm}
\centerline{\includegraphics[width=3.6in]{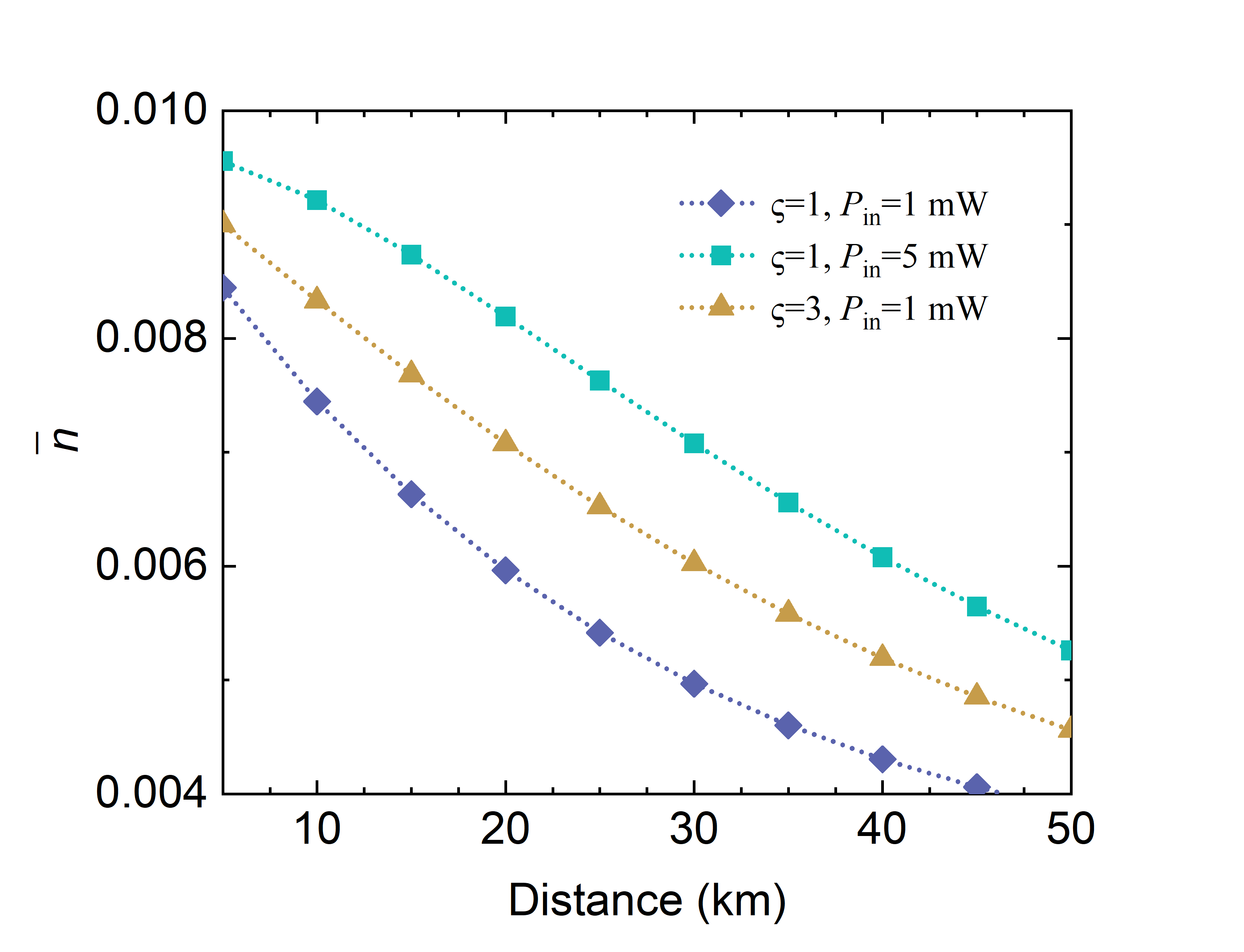}}
	\caption{\label{fig:n} The mean number of total thermal photons as a function of the transmission distance.}
\end{figure}

\begin{table}[h]
	\centering
	\caption{Physical parameters.}
	\label{tab:phy}
	\begin{tabular}{|l|c|c|}
	\hline
	{\bf Physical parameter} & {\bf Symbol} & {\bf Value} \\
	\hline
	Detection efficiency  & $ {\eta_{\rm eff}} $  &  0.7 (1.55 dB) \\
	Detection bandwidth  & W  &  100 MHz \\
	Noise equivalent power  & NEP  &  6 pW/$ \sqrt {\mathrm{Hz}} $ \\
	Detector shot-noise  & $ v_{\rm{det}} $  &  2 (het) \\
	Laser linewidth  & $ l_{\rm{sig}}, l_{\rm{LO}} $  &  1.6 kHz \\
	LO power  & $ P_{\rm{LO}} $  &  100 mW \\
	Clock  & $ \Theta $  &  5 MHz \\
	Pulse duration  & $ \delta t $  &  10 ns \\
 	RIN & $ {\Delta_{{\rm sig}}}, {\Delta_{{\rm LO}}} $ & $ 8 \times {10^{ - 11}} \mathrm{Hz}^{ - 1} $ \\
	Electric amplification & $g_e$ & 20 $ {\rm{k\Omega }} $ \\
	DAC voltage deviation & $\delta {u}$ & 0.01 $ {u_{{\rm{DAC}}}} $  \\
	CMRR & $ g_{\mathrm{C}} $ & 30 dB \\
 	PD diodes responsivity & $ \varpi $ & 20 $ {\rm{k}}\Omega $ \\
	ADC full voltage range & $\bar U$ & 1 V \\
	 Intrinsic voltage variance & $ {{V_{{\rm{intr}}}}} $ & $ {10^{ - 8}}{{\rm{V}}^2} $ \\
	ADC bit resolution & $\bar d$ & 12 \\
	\hline
	\end{tabular}
\end{table}

In what follows, we perform the numerical analysis of the total amount of thermal photons.
Fig.~\ref{fig:n} shows The mean number of total thermal photons i.e., Eq.~\eqref{e:noise} as a function of the transmission distance $L$ under various classical power and channel numbers.
The values of most of the physical parameters are assumed to be the same as Ref.~\cite{pirandola2021composable}, given in Table~\ref{tab:phy}.
The main difference is that the quantum wavelength changes from 800 nm to 1490 nm, whose fiber attenuation coefficient is 0.22 dB/km.
In Fig.~\ref{fig:n}, we find that the mean number of total thermal photons decreases as distance increases.
As discussed above, only Raman noise and phase noise are related to the transmission distance in LLO protocol.
In detail, the mean number of extra thermal photons caused by the phase noise decreases with increasing distance [see Eq.~\eqref{e:noise}], while Raman noise decreases with increasing distance after 22 km, as shown in Fig.~\ref{fig:Raman}.
In addition, the arriving thermal photons from the transmitter side i.e., $ \tau {\bar n_{{\rm{Tx}}}}$ also decreases with increasing distance.

\begin{figure}
\vspace{-0.4cm}
\centerline{\includegraphics[width=3.6in]{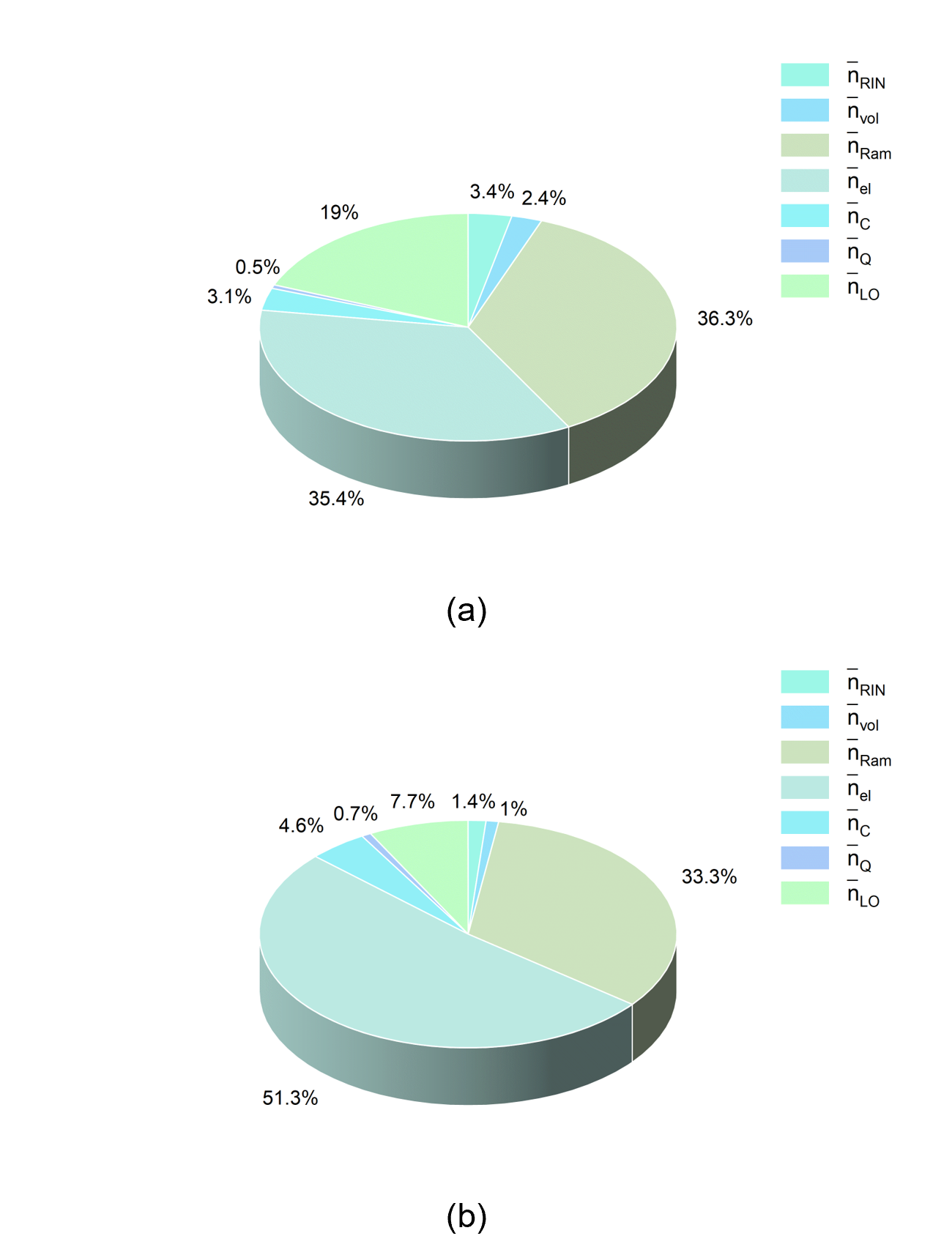}}
	\caption{\label{fig:denoise} Decomposition of the mean number of total thermal, noise photons for transmission distance (a) $L$=25 km and (b) $L$=50 km. 
    We plot the case of 1 classical channel with a 5 mW input power, while the spontaneous Raman scattering coefficient is ${\rho_\lambda }{\rm{ = 4}} \times {\rm{1}}{{\rm{0}}^{{\rm{ - 9}}}}$ ${{\rm{nm}}^{ - 1}} {{\rm{km}}^{ - 1}}$.
	}
\end{figure}

To further analyze the total amount of thermal photons, we decompose the mean number of total thermal photons in different transmission distances.
See Fig.~\ref{fig:denoise}, where we plot the case of 1 classical channel with a 5 mW input power, which corresponds to the green line in Fig.~\ref{fig:n}.
In Fig.~\ref{fig:denoise}(a), we find that phase noise, Raman noise, and electronic noise are the main noise sources, while the Raman noise is the largest for a 25 km transmission distance.
However, when the distance increases to 50 km, the electronic noise becomes the largest and accounts for more than half, as shown in Fig.~\ref{fig:denoise}(b).
Note that the electronic noise is not related to the transmission distance, which means that its value is the same at 25 km and 50 km.
Therefore, this result can be attributed to the decreased Raman noise with increased distance (see Fig.~\ref{fig:Raman}).
Moreover, the main noise sources at 50 km are still phase noise, Raman noise, and electronic noise.

\subsection{Numerical investigations} 

\begin{table}[h]
	\centering
	\caption{Composable parameters.}
	\label{tab:com}
	\begin{tabular}{|l|c|c|}
	\hline
	Parameter & Symbol & Value \\
	\hline
	PE signals & $ {{m_{{\rm{pe}}}}} $ & $0.1 \times N$ \\	Digitalization & $ d $ & $ 2^5 $ \\
	EC success probability & $ p_{\rm{ec}} $ & 0.9 \\
	$\varepsilon$-correctness & $ {{\varepsilon_{{\rm{cor}}}}} $ & $ 10^{-10} $ \\
	Smoothing parameter & $ {{\varepsilon_{{\rm{s}}}}} $ & $ 10^{-10} $ \\
	Hash parameter & $ {{\varepsilon_{{\rm{h}}}}} $ & $ 10^{-10} $ \\
	\hline
	\end{tabular}
\end{table}	

\begin{figure}
\vspace{-0.1cm}
\centerline{\includegraphics[width=3.1in]{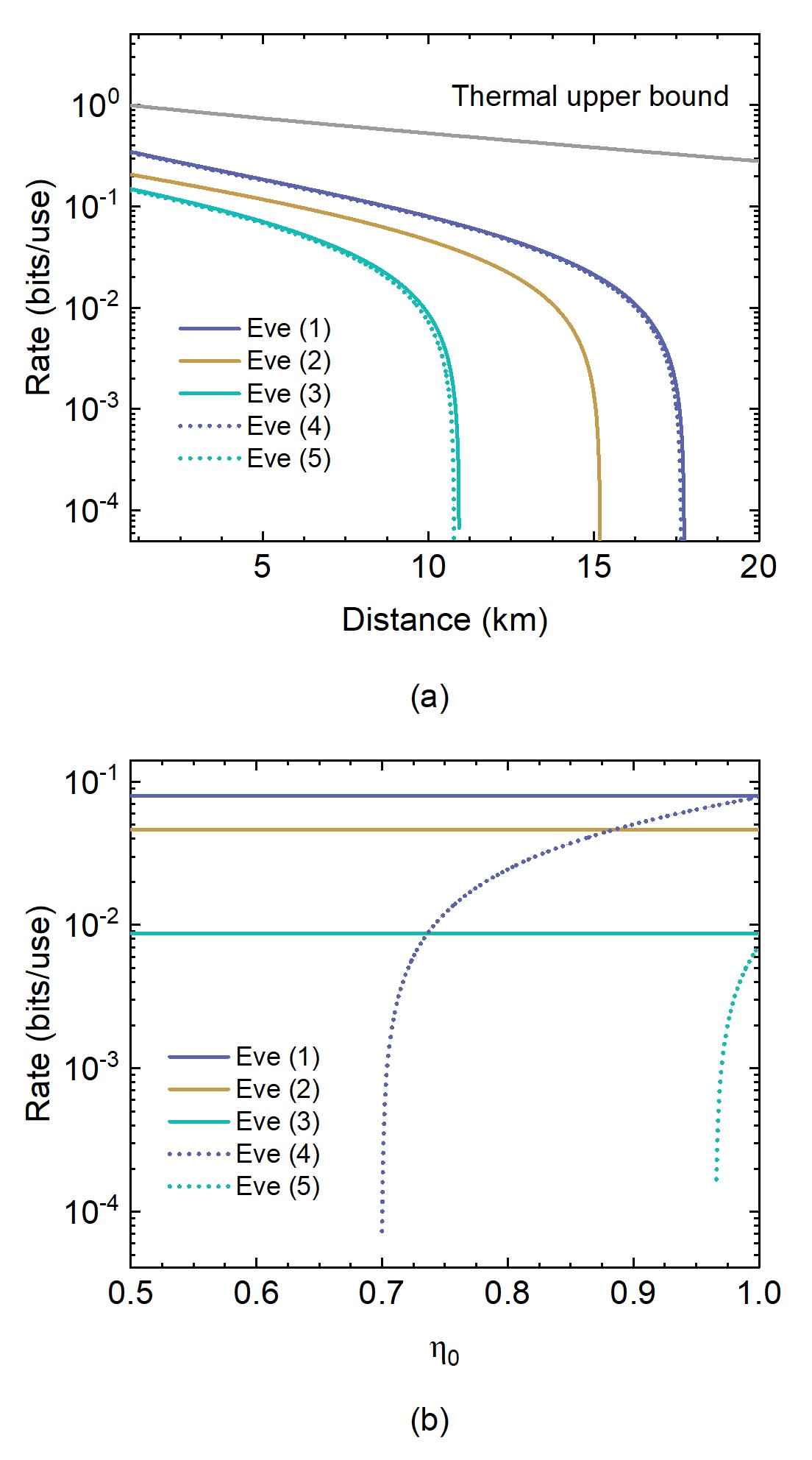}}
	\caption{\label{fig:com} 
    (a) Composable secret key rate versus transmission distance using LLO scheme and heterodyne detection under various trust levels.
    The grey line denotes the thermal-loss version of the PLOB bound~\cite{pirandola2017fundamental}.
    (b) Composable secret key rate versus $\eta_0$ under various trust levels with $L$=10 km.
    Here the classical system is set to 1 classical channel with a 5 mW input power, while the spontaneous Raman scattering coefficient is ${\rho_\lambda }{\rm{ = 4}} \times {\rm{1}}{{\rm{0}}^{{\rm{ - 9}}}}$ ${\left( {{\rm{nm}} \cdot {\rm{km}}} \right)^{ - 1}}$.
	}
\end{figure}

We plot the results in Fig.~\ref{fig:com} showing the composable finite-size key rate using LLO protocol with heterodyne detection under various trust levels.
The related composable finite-size parameters are shown in Table \ref{tab:com}.
Here the classical system has 1 classical channel with a 5 mW input power, and the finite round is $N{\rm{ = 1}}{{\rm{0}}^{\rm{7}}}$.
In Fig.~\ref{fig:com}(a), we first show a special case of $\eta_0=1$, which means Eve only controls Alice's setup noise but cannot collect leakage from Alice's setup in both Eve (4) and Eve (5) scenario.
In addition, we also plot the thermal-loss version of the PLOB bound for comparison, which is given by~\cite{pirandola2017fundamental}
\begin{equation}
\Phi \left( {\tau ,\bar n} \right) =  - {\log_2}\left[ {\left( {1 - \tau } \right){\tau ^{\frac{{\bar n}}{{1 - \tau }}}}} \right] - \iota \left( {\frac{{\bar n}}{{1 - \tau }}} \right),
\end{equation}
where $ \iota \left( \Lambda  \right): = \left( {\Lambda  + 1} \right){\log_2}\left( {\Lambda  + 1} \right) - \Lambda {\log_2}\Lambda  $.
The above equation is based on the fact that the relative entropy of entanglement suitably computed over the asymptotic Choi matrix of the thermal-loss channel provides an upper bound for its secret key capacity \cite{pirandola2017fundamental}.
Therefore, any key rate under a thermal-loss channel cannot exceed this thermal upper bound.
We see from  Fig.~\ref{fig:com}(a) that, as expected, the rate has a nontrivial improvement as a result of the stronger trust level made for the transmitter and receiver. 

In detail, the best trust level i.e., Eve (1) has a maximum safe transmission distance of 17.7 km, while the distance of the worst case i.e., Eve (5) is 10.7 km.
Note that these distances are shorter than that in recent LLO experiments \cite{jain2022practical,wang2022continuous,hajomer2024long}. 
The reason can be attributed to (i) these experiments use 1550 nm laser, whose fiber attenuation coefficient is lower, (ii) only quantum signals are propagated through a fiber so that no Raman noise from classical channel, (iii) the detection noise is considered trust, while the phase noise is lower by using advance phase compensation technology (such as machine learning method \cite{chin2021machine}), (iv) the blcok size is larger, while we set $N{\rm{ = 1}}{{\rm{0}}^{\rm{7}}}$.
In addition, we find that whether Alice's setup noise is trusted has little effect on the key rate compared to Bob's setup noise.
Therefore, the line of Eve (4) [or Eve (5)] is very close to that of Eve (1) [or Eve (3)].

However, the trust level of source-side loss cannot be ignored, as shown in Fig.~\ref{fig:com}(b).
Here we plot the composable secret key rate versus $\eta_0$ under various trust levels with $L$=10 km.
When Eve can collect leakage from Alice's setup, the rate of both Eve (4) and Eve (5) scenarios has a nontrivial decrease as $\eta_0$ decreases.
This is because the smaller $\eta_0$ is, the more leakage can be collected by Eve.
Specifically, the rate of Eve (4) starts to fall below Eve (2) when $\eta_0<0.88$, while it falls below Eve (3) if $\eta_0<0.73$.
Moreover, the Eve (4) and Eve (5) scenarios cannot generate a secret key if $\eta_0$ below 0.7 and 0.97, respectively.
Generally, the VOA at Alice's side has an attenuation above 3 dB (i.e., $\eta_0<$0.5) in LLO experiment \cite{wang2024high,Tao}, which means Alice should keep this loss trusted to get a positive secret key rate.

\section{Satellite-based protocols} \label{sec:sat} 

In this section, we show the composable finite-size key rate of satellite-based CV-QKD under various trust levels and setup noises.
As shown in previous works \cite{liao2017satellite,liao2018satellite,chen2021integrated,li2025microsatellite}, satellite-based QKD, which connects widely separated ground stations by using quantum satellites, is a promising solution with current technology to help building a global quantum network.  
Note that one can also use quantum repeaters for a global network but the realization of mature quantum repeaters remains challenging, such as quantum memory technology, limited brightness of entanglement sources, and technological restrictions associated with distillation protocols \cite{azuma2023quantum}. 
Unlike the fiber case, satellite-based CV-QKD differs in both its noise sources and trust levels. On the one hand, the source of channel noise comes from the free-space light instated of the optical fiber, which means the thermal photons related to the channel remain unknown. 
On the other hand, a new trust level will appear since the satellite-based link can be considered as a line-of-sight (LoS) link, where Alice and Bob can employ monitoring techniques to restrict eavesdropper Eve. 

In what follows, we first show the channel noise added at the free space, and then update the trust level under the LoS link.
We model the quantum satellite as a LEO satellite with a constant orbital altitude $h$ less than 2000 km. To date, this type of orbit is the first choice to demonstrate quantum communication protocols from space because of the relative ease of reaching the orbit with multiple launcher options, and the lower exposure to the aggressive ionizing radiation affecting higher altitudes. 
In addition, The rapid round-trip time around the Earth (about 1–2 hours), combined with the wide range of possible orbital inclinations, makes it possible either to cover the entire planet within hours with a single satellite or to maintain a constant position relative to the Sun \cite{pirandola2020advances}. 
Finally, we illustrate the composable finite-size key of satellite-based CV-QKD under various trust levels.

\subsection{Noise added at free space links} 

The primary source of thermal noise in free-space links is the natural brightness within the field of view of the transmission, e.g., the sky, Sun, and Moon. 
In detail, the value of such background noise depends on the operational setting (time of the day/direction of the link), the weather conditions (cloudy or clear skies), and the features of the receiver, such as its aperture $a_R$, field of view ${\Omega_{{\rm{fov}}}}$, pulse duration $ {\delta t} $, and frequency filter ${\Delta_{\rm{fs}}}$.
For convenience, one can define the following parameter
\begin{equation}\label{e:Gamma_R}
{\Gamma_R}: = {\Delta_{\rm{fs}}}\delta t{\Omega_{{\rm{fov}}}}a_R^2.
\end{equation}
In the downlink setting, the receiver at the ground station collects the direct light from the sky.
The mean number of background thermal photons per mode is
\begin{equation}
\bar n_B^{{\rm{down}}} = H_\lambda ^{{\rm{sky}}}{\Gamma_R},
\end{equation}
where $H_\lambda ^{{\rm{sky}}}$ describes the spectral irradiance of the sky in units of photons ${{\rm{m}}^{ - 2}}{{\rm{s}}^{ - 1}}{\rm{n}}{{\rm{m}}^{ - 1}}{\rm{s}}{{\rm{r}}^{ - 1}} $, and is unique to both the time of the day and the weather conditions.
For example, at the wavelength $\lambda=$ 800 nm, the values of clear night and clear day time are ${\rm{1}}{\rm{.9}} \times {\rm{1}}{{\rm{0}}^{{\rm{13}}}}$ and ${\rm{1}}{\rm{.9}} \times {\rm{1}}{{\rm{0}}^{{\rm{16}}}}$, respectively. 
In contrast to the downlink with direct light, the receiver in the uplink case is located at the quantum satellite, where the primary source of background light is the reflected sunlight injected into the aperture area.
In detail, the mean number of background thermal photons per mode is
\begin{equation}
\bar n_B^{{\rm{up}}} = \kappa H_\lambda ^{{\rm{sun}}}{\Gamma_R},
\end{equation}
where $\kappa$ is the dimensionless parameter depends on the geometry and albedos of the Earth and the Moon, 
$ H_\lambda ^{{\rm{sun}}} $ is the solar spectral irradiance in units of photons ${{\rm{m}}^{ - 2}}{{\rm{s}}^{ - 1}}{\rm{n}}{{\rm{m}}^{ - 1}}{\rm{s}}{{\rm{r}}^{ - 1}} $.

Generally, the background noise due to sunlight poses a severe limitation on the achievable performance of day-time free-space DV-QKD, limiting a lot of the demonstrations obtained so far to night-time \cite{avesani2021full}.
Fortunately, for the CV protocol, thanks to the LO acting as a noise filter, only thermal noise mode-matching with the LO will survive in the output of the CV protocol (i.e., all other noise will be filtered out), ensuring its feasibility of daytime operation \cite{pirandola2021limits}.
In detail, as shown in Eq.~\eqref{e:Gamma_R}, the number of background thermal photons per mode has a strong dependence on the frequency filter ${\Delta_{\rm{fs}}}$.
The physical frequency filters of DV protocol are typically limited to around ${\Delta_{\rm{fs}}}$= 1 nm, which corresponds to a bandwidth of $\Delta v = c{\lambda ^{ - 2}}\delta t \simeq 470$ GHz.
However, in CV-QKD, the bandwidth of the LO can be made very narrow by the current lasers so that very small values of ${\Delta_{\rm{fs}}}$ are indeed accessible.
For a pulse duration $ {\delta t} $=10 ns, one can consider ${\Delta_{\rm{fs}}}$= 0.1 pm around 800 nm, which is ${\rm{1}}{{\rm{0}}^{{\rm{ - 4}}}}$ narrower than that of DV case.
With such a filter, one has a corresponding ${\rm{1}}{{\rm{0}}^{{\rm{ - 4}}}}$ suppression of background thermal noise.
We assume a 800 nm wavelength, while the aperture and field of view are set to $a_R$=0.4 m and ${\Omega_{{\rm{fov}}}} = {10^{ - 10}}$ sr, respectively.
With a 0.1 pm frequency filter, the parameter ${\Gamma_R}$ is ${\rm{1}}{\rm{.6}} \times {\rm{1}}{{\rm{0}}^{ - 23}}{{\rm{m}}^2}$ s nm sr so that $\bar n_B^{{\rm{down}}} = 3.04 \times {10^{ - 10}}$ if the satellite operate with the downlink setting at clear night time.

\subsection{Eve (0): passive LoS security} 

One of the distinctive features of a satellite-based channel, as compared to a fiber link, is that it is a LoS link. 
While it may not be possible to fully monitor the channel between Alice and Bob, one can employ monitoring techniques, such as light detection and ranging, to detect objects of a certain minimum size along the path.
In fact, the same system and the corresponding optics that are being used for tracking and acquisition purposes can also be used to detect unwanted objects along the beam.
These techniques could restrict Eve from implementing her attack because she may find it hard to collect all signals transmitted by Alice or replace the channel between herself and Bob with a limited size.
Our previous work has divided the restricted eavesdropping into three specific setting, and proposed a so-called {\em bypass} model for analysis tool \cite{ghalaii2023satellite}.

In this study, we consider the third, most restricted setup in Ref.~\cite{ghalaii2023satellite}, where Eve is simply a passive receiver of Alice’s signal without sending anything to Bob. 
In this case, all signals received by Bob come from the bypass channel, which is inaccessible to Eve.
This is therefore a passive attack that can be interpreted as the action of a pure-loss channel, i.e., a BS with no injection of thermal photons. 
Such a passive attack plays a nontrivial role in the achievable key rate of satellite-based CV-QKD, since Eve is difficult to carry out an active attack in practice, even if no monitoring techniques are used. This is because the LEO satellite is moving all the time with a relatively high speed relative to the ground, which is a challenge for Eve to keep track. 
Under the passive LoS security, we further assume that all types of noise and loss at the setup are trusted, i.e., Eve (1).
Then, we define this trust level as Eve (0), which provides a reference for the best rate in practice.
The secret key rates under Eve (0) are derived by excluding Eve from the control of the environmental noise.
This means that her covariance matrix is reduced from that of Eve (1) to only modes $B$ and $E$, which is given by
\begin{align}
{{\rm{V}}_{BE}} = \left( {\begin{array}{*{20}{c}}
	{b{\rm{I}}}&{\theta {\rm{I}}}\\
	{\theta {\rm{I}}}&{\phi {\rm{I}}}
	\end{array}} \right).
\end{align}
The elements of the above covariance matrix are the same as that of Eve (1) [see Appendix~\ref{app:CM}].
Next, the Holevo bound for the homodyne and heterodyne settings are given by
\begin{align}
{\rm{V}}_{E|B}^{\hom } = \left( {\begin{array}{*{20}{c}}
	{\phi  - \frac{{{\theta ^2}}}{b}}&0\\
	0&\phi 
	\end{array}} \right),
\end{align}
\begin{align}
{\rm{V}}_{E|B}^{{\rm{het}}} = \left( {\phi  - \frac{{{\theta ^2}}}{{b + 1}}} \right){\rm I}.
\end{align}
Using the above Holevo bound, Eq.~\eqref{e:I} in Eq.~\eqref{e:R}, and  Eq.~\eqref{e:R_pe_com}, we compute the composable finite-size key of Eve (0) scenario.

\subsection{Numerical investigations} 

In this section, we illustrate the composable finite-size key of satellite-based CV-QKD under various trust levels.
We consider the LLO scheme using heterodyne detection, and the wavelength is set to 800 nm, which allows for a good atmospheric transmission.
Moreover, we consider the downlink setting as it shows better performance than uplink with the same atmospheric parameters.
This is because atmospheric turbulence acts in the early stages of optical signal transmission, leading to a larger diffraction-induced broadening and more serious beam wandering \cite{zuo2020atmospheric}.
However, the uplink CV-QKD is also of great value owing to the simple satellite structure, the variability of quantum sources, and the accessibility for maintenance repair.

\begin{figure}
\centerline{\includegraphics[width=3.4in]{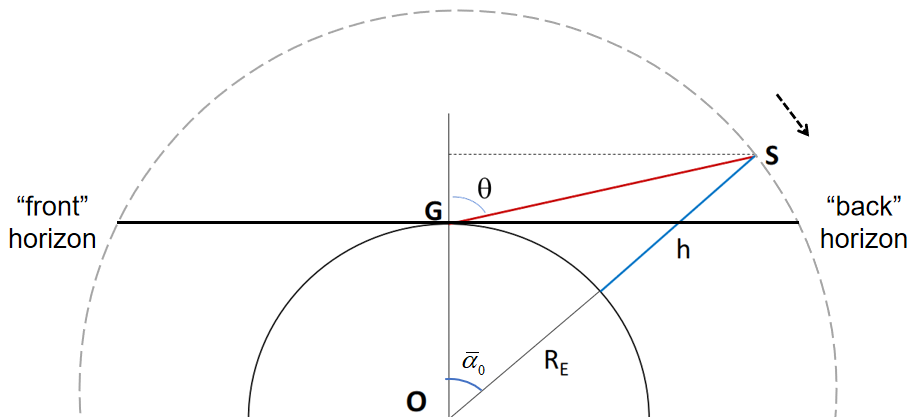}}
	\caption{\label{fig:zend} Basic geometry for satellite-based quantum communications. 
 $R_{E}$: earth radius;
 G: sea-level ground station;
 S: quantum satellite;
 $h$: satellite altitude;
 O: Earth's center;
 ${\bar \alpha_0}$: orbital angle. 
 $\theta$: zenith angle.
	}
\end{figure}

In detail, the zenith angle of the downlink setting is set to the one-radiant sector, i.e., the zenith angle $\theta  \in \left[ {{\rm{ - 1,1}}} \right]$.
Note that the zenith angle describes the angle formed between the zenith point at the ground station and the direction of observation towards the satellite, as shown in Fig.~\ref{fig:zend}.
In fact, the satellite may connect the ground station within a broader range, which arises from the “front” horizon and moves towards the zenith, and finally the “back” horizon, i.e., the zenith angle $\theta  \in \left( {{\rm{ - }}\pi /2,\pi {\rm{/2}}} \right)$.
Here, we consider the good sector $\theta \in \left[ {{\rm{ - 1,1}}} \right]$ because a larger zenith angle has a longer slant distance, so as a lower average rate.
In practice, the sector beyond $\theta  \in \left[ {{\rm{ - 1,1}}} \right]$ can be used for data processing and key generation, or encrypted communication using a previously generated key (e.g., the satellite may download the key exchanged with another station).

\begin{figure}
\centerline{\includegraphics[width=3.1in]{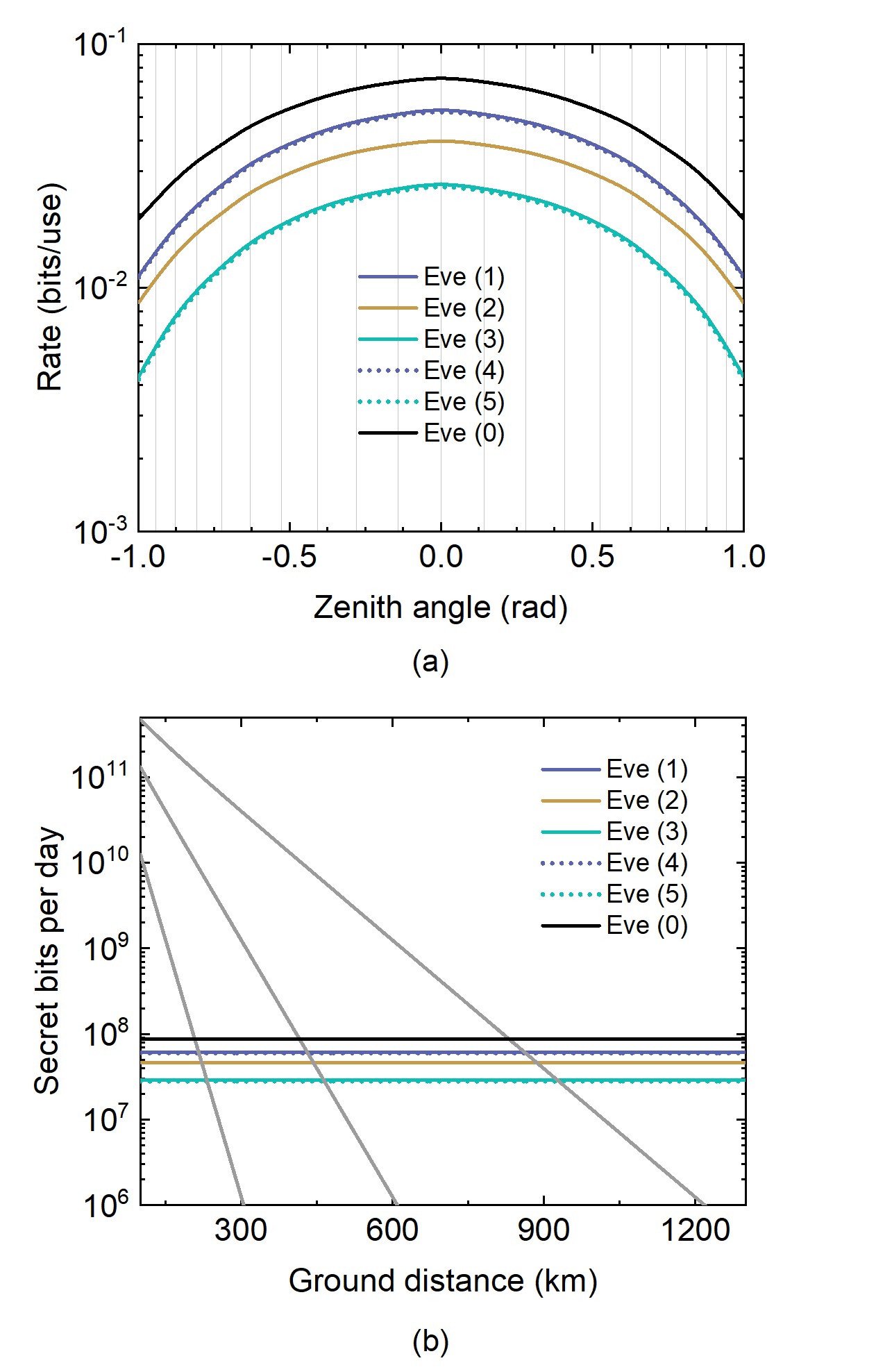}}
	\caption{\label{fig:free} 
    (a) Composable secret key rate versus zenith angle using LLO scheme and heterodyne detection under various trust levels.
    The quantum satellite, which is located at a 530 km orbit, operates at clear night time under weak turbulence.
    The grey lines denote the zenith angle ranges of the 20 slices in the one-radiant sector.
    (b) Secret-key bits per day versus ground distance between two stations, assuming a clock of 10 MHz.
    The grey lines denote the ground-based case with fiber attenuation coefficient 0.2 dB/km.
	}
\end{figure}

Fig.~\ref{fig:free} shows the composable finite-size key rate of satellite-based CV-QKD under various trust levels.
In the satellite-based case, one needs to consider the effect of the instantaneous zenith angle between the moving satellite and the ground station. This is a challenge for space systems, as the data at one block may come from various links with different zenith angles.
This leads to a problem in the theoretical analysis, as it is difficult to provide a formula describing the irregular probability distribution of transmittance during one block time.
When analyzing asymptotic security, one can simplify that the satellite is ‘frozen’ at some fixed altitude and zenith angle to avoid this issue.
This is because the asymptotic case can use the quantum channel an infinite round so has an infinite block size at each zenith angle.
However, the block size is finite in the composable finite-size case.
For this reason, we now consider a simplified but more accurate treatment where we explicitly account for the fact that different blocks of data correspond to different slices of the orbit within the one-radiant sector. 
For each slice, we consider the corresponding minimum rate, which is achieved at the largest zenith angle along that particular slice. Within this slicing model, the rate calculated in our manuscript is lower than the actual rate in operation, or in other words, it provides a pessimistic but reliable lower bound. The overall orbital rate ${R_{{\rm{orb}}}}$ is given by an average over the slices.
The details of orbital slicing are shown in Appendix~\ref{app:orb}.

Fig.~\ref{fig:free}(a) shows the composable secret key rate versus the zenith angle under various trust levels.
Here the satellite altitude is fixed at $h=530$ km (note that the altitude of the Micius satellite is between 488 and 584 km), while the operation is under weak turbulence at clear night time.
Based on Appendix~\ref{app:orb}, the transmission time of a one-radiant sector for such an orbit is ${t_Q} \simeq 200$ seconds.
Therefore, we divide the one-radiant sector into 20 slices so that each slice has a time widow $10$ s.
Specifically, the zenith angle ranges of the 20 slices are shown in the grey lines in Fig.~\ref{fig:free} (a).
For a 10 MHz clock, the block size of each slice is $N=10^8$.
Then, except wavelength $\lambda$ and the block size $N$, all other physical and composable parameters are the same as Table~\ref{tab:phy} and Table~\ref{tab:com}.
In addition, the transmittance derivation of the weak turbulence downlink is based on Ref.~\cite{pirandola2021satellite}.
In detail, the transmittance is affected by diffraction, atmospheric extinction, turbulence, and, pointing errors.
In this work, we consider that Alice employs a collimated beam (i.e., beam curvature ${R_{\rm{0}}}{\rm{ = }}\infty $), while both the initial beam spot size and receiver aperture $a_R$ are set to 0.4 m.
We find that Eve (0) with LoS security has the best rate with an overall orbital rate $R_{{\rm{orb}}}^{{\rm{Eve (0)}}}{\rm{ = }}$0.0436 bit/use.
Note that the transmittance of VOA at Alice's side is set to $\eta_0=1$ so that Eve only controls Alice's setup noise but cannot collect leakage from Alice's setup in both Eve (4) and Eve (5) scenarios.

Fig.~\ref{fig:free}(b) compares the secret-key bits per day of the satellite-based and ground-based case.
For satellite-based case, we assume that the two gorund stations has similar operational conditions and lie along the orbital path such that the satellite crosses both of their zenith positions multiple times per day.
Moreover, we consider a specific case that the two stations communicate via a satellite once a day (the satellite may be used for communication between other ground stations during the remaining time).
Accounting for the time of the quantum communication (200 s) and 10 MHz clock, the secret bits distributed for each night-time zenith-crossing passage is
\begin{equation}
R_{{\rm{daily}}}^{{\rm{sat}}} \approx \Theta  \cdot {t_Q} \cdot {R_{{\rm{orb}}}},
\end{equation}
For example, the secret-key bits of Eve (0) scenario per day is $ \simeq {\rm{8}}{\rm{.72}} \times {\rm{1}}{{\rm{0}}^{\rm{7}}}$ secret bits.
For ground-based case, we consider the maximum number of secret bits per day (versus ground distance) that can be distributed by a repeaterless fiber link between the stations and also by fiber links assisted by ideal quantum repeaters. 
Assuming the same clock $\Theta$=10 MHz, we compare these ground-based performances with the secret-bits per day that can be distributed by using a satellite moving between the two stations.
In detail, the fiber-based rate with ideal quantum repeaters is \cite{pirandola2019end}
\begin{equation}
R_{{\rm{fib}}}^{{\rm{rep}}} =  - {\log_2}\left( {1 - \sqrt[{{N_{{\rm{rep}}}} + 1}]{{{\eta_{{\rm{fib}}}}}}} \right),
\end{equation}
with the repeater number ${{N_{{\rm{rep}}}}}$.
Then the corresponding number of secret bits per day is 
\begin{equation}
R_{{\rm{daily}}}^{{\rm{fib}}} \approx \Theta  \cdot R_{{\rm{fib}}}^{{\rm{rep}}} \cdot T,   
\end{equation}
where $T \simeq {\rm{8}}.{\rm{6}} \times {\rm{1}}{{\rm{0}}^4}$ s.
Based on Fig.~\ref{fig:free}(b), we find that how a satellite can beat the fiber-based repeaterless bound, and how it can achieve the same rate with the fiber links assisted by ideal quantum repeaters.
Here the fiber-based rate is calculated using a 1550 nm wavelength with attenuation coefficient 0.2 dB/km, which is lower than that of Sec.~\ref{sec:fib}.
For example, in the worst-case Eve (5), a satellite can beat the fiber-based repeaterless bound when the stations are separated by more than 233 km, and it can achieve the same rate of 3 ideal quantum repeaters when the station-to-station ground separation is about 930 km. 
Note that the composable finite-size of satellite-based cased can be further improved by the postselection or clusterization technology \cite{usenko2012entanglement,ruppert2019fading}.

\section{Conclusion} \label{sec:con}

In this paper, we have investigated the composable secret key rates that are achieved under various trust levels, from the worst-case assumption of a fully untrusted setup to the case where the setup’s loss and noise are considered to be trusted.
We have shown how the realistic assumptions on the setup (both transmitter and receiver) can have nontrivial effects in terms of increasing the composable key rate and tolerating higher loss. 
More interestingly, we have also demonstrated that a practical one-way GMCS protocol (with an LLO scheme) is hard to be robust against the untrusted loss on Alice's side, which highlights the importance of source-side security or isolation.

Besides the general fiber-based case, we have also analyzed the satellite-to-ground communication with an LEO quantum satellite, considering the downlink setting with night time operation and the effect of the instantaneous zenith angle.
The unique trust level of the satellite-based case with LoS security is also considered.
In this way, we have discussed that its key distribution rate is able to outperform a ground chain of ideal quantum repeaters even with the worst trust level, showing the potentiality of satellite for building a high-rate and large-scale CV network with current technology.
 
\section*{Acknowledgments}

\noindent Z.Z. and S.P. thank Fabian Laudenbach for comments.
Z.Z. also thanks Tao Wang and Yuehan Xu for helpful discussion on experiment parameters. S.P. acknowledges support from UKRI via the ``Integrated Quantum Networks Research Hub'' (IQN, EP/Z533208/1).
Z.Z. is supported by Quantum Science and Technology-National Science and Technology Major Project (Grant No. 2021ZD0300703).

\appendix 

\section{The elements of $ C $ and $ {{\rm{V}}_{EE''}} $ under Eve (1) to Eve (4)} \label{app:CM}

For Eve (1) scenario, Eve only controls the quantum channel so that Eve's modes have a variance $ \omega  = 2{\bar n_{{\rm{Eve}}(1)}} + 1 = 2{\bar n_e} + 1 $ and then input the channel with the transmittance $1-\eta_{\rm ch}$.
Based on a similar derivation in Sec.~\ref{sec:ele}, the remaining elements of $ C $ and $ {{\rm{V}}_{EE''}} $ have the form
\begin{align}\label{e:1_1}
\psi  = \sqrt {{\eta_{{\rm{ch}}}}\left( {{\omega ^2} - 1} \right)},
\end{align}
\begin{align}
\quad \phi  = {\eta_{{\rm{ch}}}}\omega  + \left( {1 - {\eta_{{\rm{ch}}}}} \right)\left( {\mu  + 2{{\bar n}_{{\rm{Tx}}}}} \right),
\end{align}
\begin{align}
\gamma  = \sqrt {{\eta_{{\rm{eff}}}}\left( {1 - {\eta_{{\rm{ch}}}}} \right)\left( {{\omega ^2} - 1} \right)}, 
\end{align}
\begin{align}
\quad \theta  = \sqrt {\tau \left( {1 - {\eta_{{\rm{ch}}}}} \right)} \left( {\omega  - \mu  - 2{{\bar n}_{{\rm{Tx}}}}} \right).
\end{align}
For Eve (2) with untrusted-loss and trusted-noise detector, the variance of Eve's mode becomes $ \omega  = 2{\bar n_{{\rm{Eve}}(2)}} + 1 $, while its input transmittance is modified to $1-\tau$.
Then, the elements are
\begin{align}\label{e:Eve2:1}
\psi  = \sqrt {\tau \left( {{\omega ^2} - 1} \right)}, 
\end{align}
\begin{align}
\quad \phi  = \tau \omega  + \left( {1 - \tau } \right)\left( {\mu  + 2{{\bar n}_{{\rm{Tx}}}}} \right),
\end{align}
\begin{align}
\gamma  = \sqrt {\left( {1 - \tau } \right)\left( {{\omega ^2} - 1} \right)} ,
\end{align}
\begin{align}\label{e:Eve2:4}
\quad \theta  = \sqrt {\tau \left( {1 - \tau } \right)} \left( {\omega  - \mu  - 2{{\bar n}_{{\rm{Tx}}}}} \right).
\end{align}
For Eve (3) with an untrusted detector, Eve's modes have a variance $ \omega  = 2{\bar n_{{\rm{Eve}}(3)}} + 1 $ with the same input transmittance as Eve (2).
The derivation shows that except $\omega$, the remain elements are the same as those of Eve (2), i.e., Eq.~\eqref{e:Eve2:1} to Eq.~\eqref{e:Eve2:4}.
For Eve (4) with untrusted Alice, the variance of Eve's modes become $ \omega  = 2{\bar n_{{\rm{Eve}}(4)}} + 1 $, and the transmittance is modified to $1-{\eta_{\rm{0}}}{\eta_{{\rm{ch}}}}$.
Finally, the remaining elements of $ C $ and $ {{\rm{V}}_{EE'}} $ are
\begin{align}
\psi  = \sqrt {\tau '\left( {{\omega ^2} - 1} \right)},
\end{align}
\begin{align}
\phi  = \tau '\omega  + \left( {1 - \tau '} \right){\mu_0},
\end{align}
\begin{align}
\gamma  = \sqrt {{\eta_{{\rm{eff}}}}\left( {1 - \tau '} \right)\left( {{\omega ^2} - 1} \right)},
\end{align}
\begin{align}
\quad \theta  = \sqrt {{\tau_0}\left( {1 - \tau '} \right)} \left( {\omega  - {\mu_0}} \right).
\end{align}
Note that the details of ${\bar n_{{\rm{Eve}}(2)}}$, ${\bar n_{{\rm{Eve}}(3)}}$, and ${\bar n_{{\rm{Eve}}(4)}}$ can be found in Sec.~\ref{sec:ele}.

\section{Derivation of voltage deviation noise}\label{app:mod}

This Appendix first show the derivation of voltage deviation noise on the quadrature operator $ \hat q $, which is related to MZM1 in Fig.~\ref{fig:system}(b).
In the ideal case, the output of MZM1 is given by
\begin{equation}
{\alpha_I} = \frac{1}{2}{\alpha_{{\rm{in}}}}\cos \left( {\frac{{{u_I} + {u_{0,I}}}}{{2{u_\pi }}}\pi } \right).
\end{equation}
Next, this output is attenuated by the VOA in Fig.~\ref{fig:system}(a) to
\begin{equation}\label{e:att1}
{\alpha_{I,{\rm{att}}}} = \frac{{\sqrt {{\eta_0}} }}{2}{\alpha_{{\rm{in}}}}\cos \left( {\frac{{{u_I} + {u_{0,I}}}}{{2{u_\pi }}}\pi } \right): = \frac{q}{2},
\end{equation}
which is considered as the amplitude of the position component of the coherent state $ \left| \alpha  \right\rangle $.
If we consider the deviation of both $ {{u_I}} $ and $ {{u_{0,I}}} $, the non-ideal output becomes
\begin{equation}
{{\bar \alpha }_{I,{\rm{att}}}} = \frac{{\sqrt {{\eta_0}} }}{2}{\alpha_{{\rm{in}}}}\cos \left( {\frac{{{u_I} + {u_{0,I}}}}{{2{u_\pi }}}\pi  + \frac{{g\delta {u } + \delta {u_{{\rm{DC}}}}}}{{2{u_\pi }}}\pi } \right).
\end{equation}
Let us make a Taylor expansion on $ {{\bar \alpha }_{I,{\rm{att}}}} $ and then minus Eq.~\eqref{e:att1} to obtain the output derivation, which can be expressed as
\begin{equation}
\delta {\alpha_{I,{\rm{att}}}} =  - \frac{{\sqrt {{\eta_0}} }}{2}{\alpha_{{\rm{in}}}}\left[ {Y\sin \left( X \right) + \frac{{{Y^2}}}{2}\cos \left( X \right)} \right],
\end{equation}
with
\begin{equation}
X: = \frac{{{u_I} + {u_{0,I}}}}{{2{u_\pi }}}\pi, \quad Y: = \frac{{g\delta {u} + \delta {u_{{\rm{DC}}}}}}{{2{u_\pi }}}\pi.
\end{equation}
Because both $ {\sin \left( X \right)} $ and $ {\cos \left( X \right)} $ are between -1 and 1, one can get the upper bound
\begin{equation}
\left| {\delta {\alpha_{I,{\rm{att}}}}} \right| \le \frac{{\sqrt {{\eta_0}} }}{2}\left| {{\alpha_{{\rm{in}}}}} \right|\left[ {Y + \frac{{{Y^2}}}{2}} \right].
\end{equation}
Therefore, the upper bound of the injecting extra thermal photons on $ \hat q $ is given by
\begin{equation}\label{e:vol:q}
\bar n_{{\rm{vol}}}^q = {\left| {\delta {\alpha_{I,{\rm{att}}}}} \right|^2} \le \frac{{{\eta_0}}}{4}{\left| {{\alpha_{{\rm{in}}}}} \right|^2}{\left[ {Y + \frac{{{Y^2}}}{2}} \right]^2}.
\end{equation}
The last step is to determine the range of $ {\left| {{\alpha_{{\rm{in}}}}} \right|^2} $, which is the average photon number of modulator input.
After the I/Q modulator, the average photon number follows $ {\left| {{\alpha _{{\rm{out}}}}} \right|^2} \le {{{{\left| {{\alpha _{{\rm{in}}}}} \right|}^2}} \mathord{\left/
 {\vphantom {{{{\left| {{\alpha _{{\rm{in}}}}} \right|}^2}} 2}} \right.
 \kern-\nulldelimiterspace} 2}$~\cite{laudenbach2018continuous}.
Then, since the average photon number of the attenuated coherent state has $ {\left| {{\alpha }} \right|^2}=\eta_{0}{\left| {{\alpha_{{\rm{out}}}}} \right|^2} $, Eq.~\eqref{e:vol:q} can be further simplified to 
\begin{equation}\label{e:vol:q:2}
\bar n_{{\rm{vol}}}^q \le \frac{1}{2}{\left| \alpha  \right|^2}{\left[ {Y + \frac{{{Y^2}}}{2}} \right]^2}.
\end{equation}
Finally, the derivation of the voltage deviation noise on the quadrature $ \hat p $ follows the same steps and has the same value.
Therefore, the total voltage deviation noise is 2 times Eq.~\eqref{e:vol:q:2}, which is the Eq.~\eqref{e:vol} in Sec.~\ref{sec:mod}.
 
\section{Derivation of Raman scatter power}\label{app:Raman}

\begin{figure}
	\centerline{\includegraphics[width=3.6in]{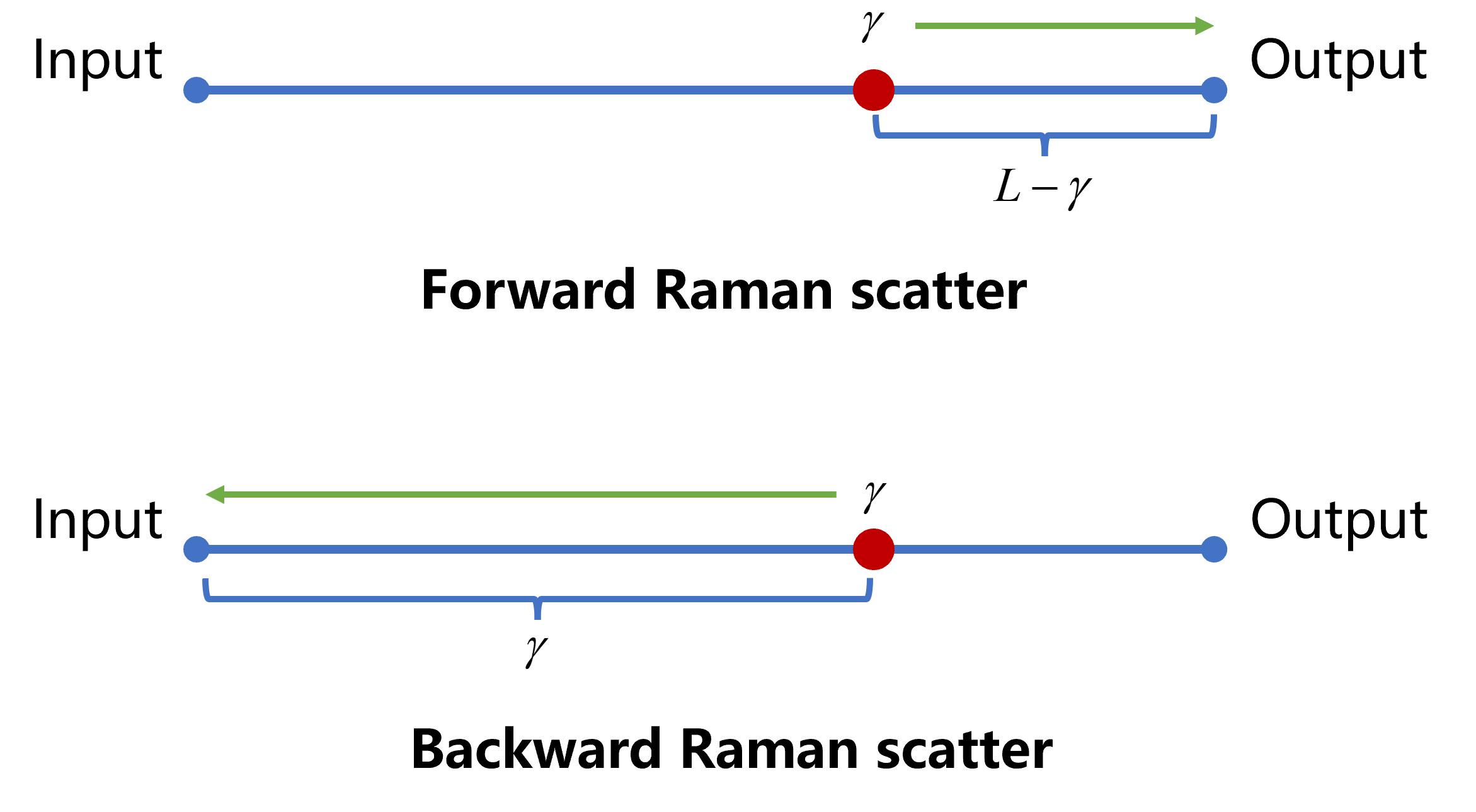}}
	\caption{\label{fig:Raman:A} The forward (up) and backward (down) Raman scatter in position $\gamma$ of the fiber.
    $L$ is the total transmission distance.
	}
\end{figure}

The Raman scatter power $ {\rm{d}}{P_{{\rm{Ram}}}} $ at wavelength $\lambda$ from a fiber element of length $ {\rm{d}}x $ at position $\gamma $ when a power $P_{\rm{in}}$ is launched into a fiber is
\begin{align}
{\rm{d}}{P_{{\rm{Ram}}}} = {P_{{\rm{in}}}} \cdot {10^{ - {\bar \alpha} \gamma /10}} \cdot {\rho_\lambda } \cdot {\Delta_{{\rm{fs}}}}\;{\rm{d}}\gamma.
\end{align}
Note that the scatter from a single fiber element $ {\rm{d}}x $ is almost isotropic. 
Therefore, we have to account for the attenuation of the fiber (length $L$) when the scatter propagates to the fiber output (forward scatter) or back to the fiber input (backward scatter), as shown in Fig.~\ref{fig:Raman:A}. 
Then, we have
\begin{align}
{\rm{d}}{P_{{\rm{Ram}},{\rm{f}}}} = {\rm{d}}{P_{{\rm{Ram}}}} \cdot {10^{ - {\bar \alpha} \left( {L - \gamma } \right)/10}},
\end{align}
\begin{align}
{\rm{d}}{P_{{\rm{Ram}},{\rm{b}}}} = {\rm{d}}{P_{{\rm{Ram}}}} \cdot {10^{ - {\bar \alpha} \gamma /10}},
\end{align}
where $ {\rm{d}}{P_{{\rm{Ram}},{\rm{f}}}} $ and $ {\rm{d}}{P_{{\rm{Ram}},{\rm{b}}}} $ denotes the case of forward and backward direction, respectively.
Integrating over the whole fiber, we have
\begin{equation}
{P_{{\rm{Ram}},{\rm{f}}}}={P_{{\rm{in}}}} \cdot {10^{ - {\bar \alpha} L/10}} \cdot L \cdot {\rho_\lambda } \cdot {\Delta_{{\rm{fs}}}}, 
\end{equation}
\begin{equation}
{P_{{\rm{Ram}},{\rm{f}}}}={\rm{5}}{P_{{\rm{in}}}} \cdot {\rho_\lambda } \cdot {\Delta_{{\rm{fs}}}} \cdot \frac{{1 - {{10}^{ - {\bar \alpha} L/5}}}}{{{\bar \alpha} \ln \left( {10} \right)}},
\end{equation}
which are shown in Eq.~\eqref{e:forward} and Eq.~\eqref {e:backward}.

~\\

\section{Orbit slicing}\label{app:orb}

Fig.~\ref{fig:zend} shows the basic geometry for satellite-based quantum communications.
According to Newton's second law, the satellite speed $v$ can be written as a function of satellite altitude $h$, which is given by
\begin{equation} 
v=\sqrt{\frac{G_0 M}{R_{E}+h}},
\end{equation}
where $G_0$ is the universal gravitation, $M=5.97\times10^{24}$ kg denotes the earth mass, and $R_{E}=6371$ km represents the earth radius.
Therefore, the orbital period (in seconds)
\begin{equation}
{T_S}{\rm{ = 2}}\pi \frac{{{R_E} + h}}{v}.
\end{equation}
As a result, the orbital angle ${\bar \alpha_0}$ varies over time $t$ according to the law
\begin{equation}
{\bar \alpha_0}\left( {t,h} \right) = \frac{{{\rm{2}}\pi t}}{{{R_E} + h}},
\end{equation}
where we have implicitly set ${\bar \alpha_0}=0$ (satellite at the zenith) for the instantaneous time $t$=0.
Correspondingly, the time-varying zenith angle can be computed from
\begin{equation}
\sin \theta  = \frac{{\left( {{R_E} + h} \right)\sin {{\bar \alpha_0}}}}{{z\left( {h,{{\bar \alpha_0}}} \right)}},
\end{equation}
with the slant distance of a zenith-crossing circular orbit given by
\begin{equation}
z\left( {h,{{\bar \alpha_0}}} \right) = \sqrt {R_E^2 + {{\left( {{R_E} + h} \right)}^2} - 2{R_E}\left( {{R_E} + h} \right)\cos {{\bar \alpha_0}}}.
\end{equation}

\bibliography{references.bib}

\end{document}